\documentclass[acmlarge]{acmart}

\usepackage[utf8]{inputenc}

\usepackage{xcolor,colortbl}
\usepackage{epstopdf} 
\usepackage{verbatim}
\usepackage{pgfplotstable,filecontents}
\pgfplotsset{compat=1.9}
\usepackage{enumitem}
\usepackage{ulem}
\usepackage{breakurl}
\usepackage{breakcites}
\PassOptionsToPackage{hyphens}{url}
\usepackage{hyperref}
\hypersetup{colorlinks=true,urlcolor=blue}

\usepackage{graphicx}
\usepackage{grffile}

\usepackage{soulutf8} 
\renewcommand\hl[1]{#1} 

\usepackage{wrapfig}
\usepackage{longtable,tabu}

\usepackage{multirow}
\usepackage{multicol}
\usepackage{booktabs}
\usepackage{threeparttable}

\usepackage{caption}
\usepackage{subcaption}
\usepackage{rotating}
\usepackage{lipsum}

\usepackage{amsmath, breqn}
\newcommand{\smalltilde}{\raise.17ex\hbox{$\scriptstyle\mathtt{\sim}$}} 

\usepackage{fancyhdr}

\setdescription{topsep=0pt, partopsep=0pt, parsep=0pt, itemsep=0pt, leftmargin=.15cm}

\usepackage{changepage}

\begin{document}

\setcopyright{acmcopyright}
\copyrightyear{2022}
\acmYear{2022}
\acmDOI{XXXXXXX.XXXXXXX}

\acmJournal{DTRAP}
\acmVolume{00}
\acmNumber{00}
\acmArticle{000}
\acmMonth{1}


\title[Beyond the Hype]{Beyond the Hype: An Evaluation of Commercially Available Machine-Learning-Based Malware Detectors}
\thanks{\hl{Code for cost model}: 
\url{https://github.com/bridgesra/beyond-the-hype-paper-code}}



\author{Robert~A.~Bridges}
\orcid{0001-7962-6329}
\email{bridgesra@ornl.gov} 

\author{Sean~Oesch}
\email{oeschts@ornl.gov} 

\author{Michael~D.~Iannacone}
\email{iannaconeme@ornl.gov} 

\author{Kelly~M.~T.~Huffer}
\email{hufferkm@ornl.gov} 

\author{Brian~Jewell}
\email{jewellbc@ornl.gov} 

\author{Jeff~A.~Nichols}
\email{nicholsja@ornl.gov}

\author{Brian~Weber}
\email{weberb@ornl.gov}
\affiliation{%
    \institution{Oak Ridge National Laboratory}
    \city{Oak Ridge} 
    \state{TN}
    \country{USA}
    }

\author{Miki~E.~Verma}
\email{meverma@stanford.edu} 
\affiliation{%
    \institution{Stanford University}
    \city{Palo Alto} 
    \state{CA}
    \country{USA}
}

\author{Daniel~Scofield}
\author{Craig~Miles}
\affiliation{
    \institution{Amazon Inc.}
    \country{USA}
}

\author{Thomas~Plummer}
\email{thomas.plummer@lmco.com}
\author{Mark~Daniell}
\email{mark.a.daniell@lmco.com}
\affiliation{
    \institution{Lockheed Martin}
    \country{USA}
}

\author{Anne~M.~Tall}
\email{annetall@mitre.org}
\affiliation{
    \institution{MITRE Corporation}
    \country{USA}
}

\author{Justin M. Beaver} 
\email{jbeaver@lirio.com}
\affiliation{
    \institution{Lirio LLC}
    \country{USA}
}

\author{Jared~M.~Smith}
\email{jaredsmith@securityscorecard.io}
\affiliation{
    \institution{Security Scorecard}
    \country{USA}
    }

\renewcommand{\shortauthors}{Bridges, et al.}

\begin{abstract}
There is a lack of scientific testing of commercially available malware detectors, especially those that boast accurate classification of never-before-seen (i.e., zero-day) files using machine learning (ML). 
Consequently, efficacy of malware detectors is opaque, inhibiting end users from making informed decisions and researchers from targeting gaps in current detectors. 
In this paper, we present a scientific evaluation of four \hl{prominent commercial} malware detection tools to assist an organization with two primary questions:  
To what extent do ML-based tools accurately classify previously and never-before-seen files? 
\hl{Is purchasing a network-level malware detector worth the cost?} 
To investigate, we tested each tool against 3,536 total files (2,554 or 72\% malicious, 982 or 28\% benign) of a variety of file types, including hundreds of malicious zero-days, polyglots, and APT-style files, delivered on multiple protocols.
We present statistical results on detection time and accuracy, consider complementary analysis (using multiple tools together), and provide two novel applications of the recent cost-benefit evaluation procedure of Iannacone \& Bridges. 
Although the ML-based tools are more effective at detecting zero-day files and executables, the signature-based tool might still be an overall better option. 
Both network-based tools provide substantial (simulated) savings when paired with either host tool, yet both show poor detection rates on protocols other than HTTP or SMTP. 
Our results show that all four tools have near-perfect precision but alarmingly low recall, especially on file types other than executables and office files---37\% of malware, including all polyglot files, were undetected. 
Priorities for researchers and takeaways for end users are given. 
\hl{Code for future use of the cost model is provided.} 
\end{abstract}

\ccsdesc[500]{Security and privacy~Malware and its mitigation}
\ccsdesc[500]{Security and privacy~Economics of security and privacy}

\keywords{malware detection, endpoint detection, network detection, evaluation, test, intrusion detection, cost benefit analysis, static analysis, dynamic analysis, machine learning}

\settopmatter{printfolios=true}
\pagenumbering{arabic}

\thanks{\hl{redacted organizational disclaimer}}

\maketitle

\section{Introduction} 
\label{sec:intro}
Attackers use malicious software, known as \textit{malware}, to steal sensitive data, damage network infrastructure, and hold information for ransom. 
One of the top priorities for computer security tools is to detect malware and prevent or minimize its impact on both corporate and personal networks. 
Traditionally, signature-based methods have been used to detect files previously identified as malicious with near perfect precision, but potentially miss newer malware samples.
With the advent of self-modifying malware and the rapid increase in novel threats, signature-based methods are insufficient on their own. 
By generalizing patterns of known benign/malicious training examples, machine learning (ML) exhibits the capability to quickly and accurately classify novel file samples in many research studies \cite{Gibert_Mateu_Planes_2020}. 
Moreover, ML-based malware research has made the transition from the 
\begin{wrapfigure}[]{r}{.45\textwidth}
    \centering
    \includegraphics[width=.45\textwidth]{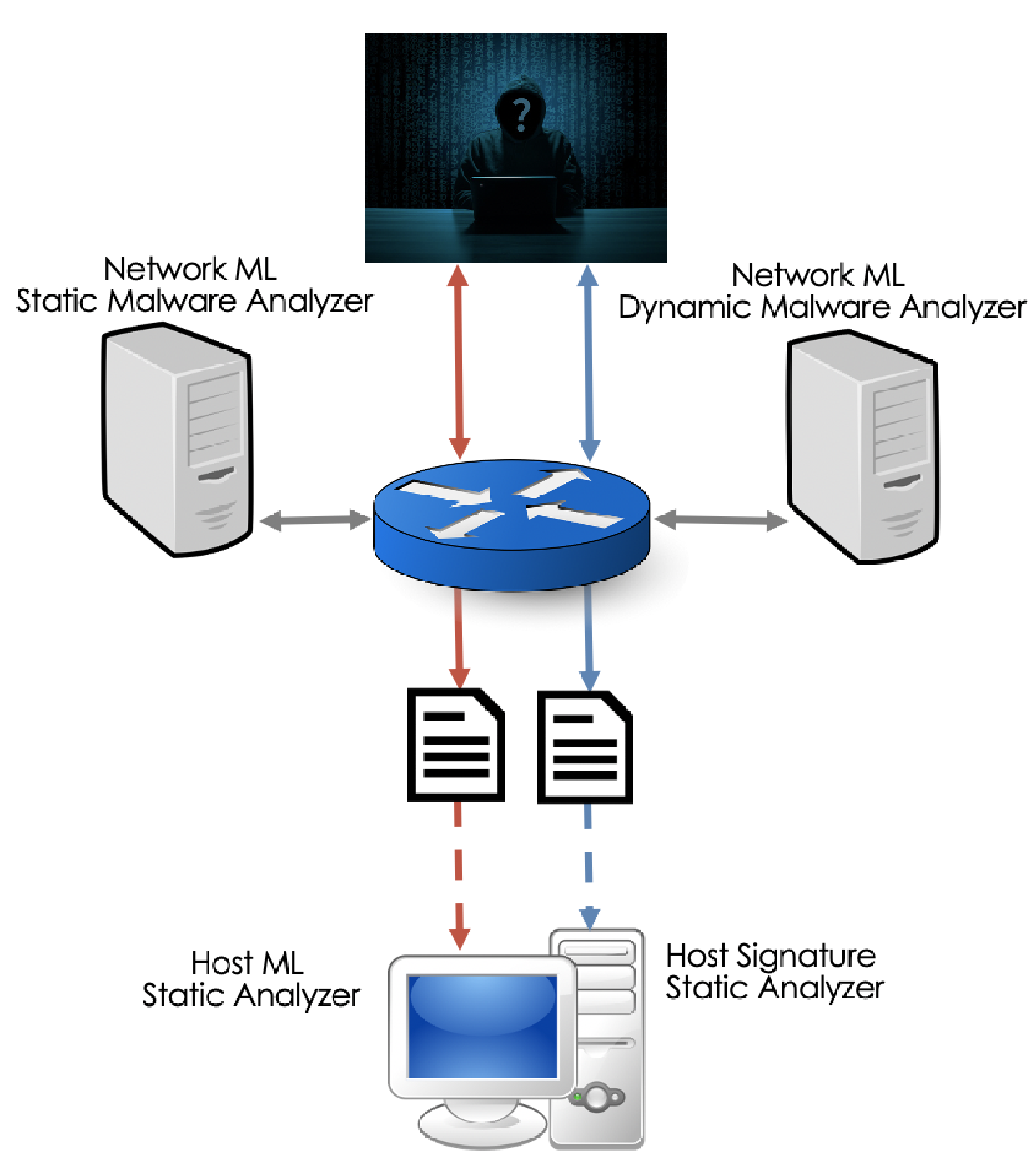}
    \caption{Malware detection experiment}
    \label{fig:experiment-diagram}
\end{wrapfigure}
subject of myriad research efforts to a current mainstay of commercial-off-the-shelf (COTS) malware detectors. 
Yet, few practical evaluations of COTS ML-based technologies have been conducted. 

\indent\rule{0pt}{0pt}\hl{
Turning from the academic literature to market reports from commercial companies can provide (for a fee) useful information, specifically, end-user feedback, itemization of all technologies in the antivirus/endpoint detection and response marketplace {\cite{gartner-edr}}, and  even statistics showing the efficacy of the detectors on malware tests {\cite{av-test, selabs-test}}. 
(See Related Works {\ref{sec:market-reports}} for recent examples by AV-TEST and SE Labs.) 
Yet, these tests often report near perfect detection for many tools, diminishing trust in the capability of the test to sufficiently stress test the tools or to differentiate the tools.
Further, as our results---under 60\% recall of all tools tested---and other previous works {\cite{christodorescu2004testing}} confirm, these near perfect evaluations promote a false narrative that modern (especially ML-driven) malware detectors provide just shy of perfect protection. 
Even when armed with many statistics on detection capabilities, balancing or weighing these heterogeneous measurements when reasoning about the efficacy of tools is difficult {\cite{iannacone2020quantifiable}}. 
}

As a result, organizations have limited insight into whether or how COTS malware detection tools add value to their current defense solutions, short of trusting possibly skewed vendor-provided claims and market/consumer reports. 
The origin of this paper is evidence of this fact, as the authors were tasked with evaluating four market-leading  malware detection tools to assist in understanding their efficacy. 
This work describes a scientific experiment designed and conducted by the authors (professional researchers) to assist an organization in examining the benefits and tradeoffs of  commercially available ML-based malware detection tools.

\hl{
Guided by the organization's requests, we established meetings with 22 sales and technical representatives of many popular vendors to learn the merits and approaches of their detection technologies. 
Based on the organization's requirements, four prominent detectors were chosen for study: two host-based (one signature-based, one ML-based) and two network-based (both ML-based, one static, and one dynamic) detectors. 
Of particular interest in this study are 
detection capability across varying file types and categories; 
efficacy of  the network-level detectors across varying protocols; and 
studying value provided by ML- vs. signature-based, host vs. network, and static vs. dynamic tools, with an eye towards ``defense-in-depth,'' (i.e.,  strategically combining tools).  
}
Our overall goal is to provide the funding organization with methods for quantitatively reasoning about malware detectors and to exhibit results on these four tools that can provide insights into two questions: 

\textbf{(Q1) ML Generalization Hypothesis:} We are interested in the fundamental promise of the three ML-based malware detectors; namely,  
``Can ML-based COTS tools accurately classify both never-before-seen files (especially zero-day malware) and publicly available files?''

\textbf{(Q2) Network-Level Malware Detection Hypothesis:} Network-level malware detectors seek to complement host detectors. They leverage the advantage of greater computational resources (available in a network appliance, generally a commodity server) without affecting users' workstations. 
\hl{ This means, in theory, they can afford more in-depth analysis without impacting usability, namely slowing processes on each workstation.
Notably, the network-level detector must carve the file out of the packets to correctly process it, which is not necessary for host detectors.} 
Of course, if the network detector cannot expand the detection capabilities, it simply adds cost without enhancing defenses. 
In our experience with approximately ten security operation centers (SOCs) 
\cite{oesch2020assessment, bridges2018how},
all centers required endpoint malware detection and signature-based network intrusion detection, but none used network-level malware detection technologies. 
Thus, we investigate the question, ``Is it worth purchasing, configuring, and maintaining a network-level malware detector, given in-place host-based detection?''


We were given licenses from four vendors for experimentation under the agreement that results could only be released with anonymity; we are unable to provide any details that might disclose the vendors or their intellectual property. 
The four COTS tools used in this experiment---two endpoint detection tools, one signature based and one claiming to be solely ML based (supervised learning with no signatures), and two network-level tools, both ML based with one claiming static analysis (does not execute files) and the other claiming dynamic analysis (executes files in a sandbox)---provide representatives of different available malware detection approaches.  
The first endpoint malware detector being solely signature based is considered a baseline to represent the status quo for modern commercial malware defenses.  
Both network detection tools perform file extraction from packets (whereas endpoint detectors scan files on each host), and both claim to use ML techniques to detect files. 
The first tool uses static analysis and an ensemble of ML- and signature-based detectors.  
The second  tool performs dynamic analysis to feed an ML classifier.
See Section \ref{sec:tools}. 
\hl{We note that two types of commercially available malware detectors are not represented in this study:  those that leverage a cloud connection for interactive intelligence (which can be at the host and/or network level) and network-level detectors that are solely signature based. }

To test these technologies, we configured a network at the National Cyber Range (NCR) \cite{ferguson2014ncr} (see Section \ref{sec:experiment}) to deliver 2,554 malware and 982 benignware samples of varying file types to hosts.  
Zero-days (specially crafted malware never seen in the wild), files to mimic advanced persistent threat (APT) actions, and polyglots (files functioning with multiple file types) were also included to test the capability of these tools to identify novel threats (see Figure \ref{fig:experiment-diagram}).

Statistical results (Section \ref{sec:stats-results}) from the experiment give enlightening insight into how well or how poorly commercially available malware detectors perform. 
Analysis of detection statistics by file type and delivery protocol illuminates rather alarming gaps. 
Comparison of the signature-based tools to the three ML tools, especially on zero-day files, allows us to quantify question (Q1), the ML generalization hypothesis. 
Furthermore, we provide an analysis of conviction latency (i.e., time to detect) per tool and per file type to illustrate the gains/losses of network tools and of using static versus dynamic analysis. 
Under the assumption that host-based malware detection is a requirement,  we consider a defense-in-depth question, ``Is it worthwhile to use multiple detectors together?,'' by simulating complementary detectors using the logical \texttt{OR} of their alerts.   

While these statistical results are enlightening, they present tradeoffs that are difficult to reconcile when choosing the ``best'' malware detection tools (e.g.,  how to balance give and take among detection rates, false alerts, detection delays, and anticipated attack costs). 
To aid the funding organization in using all the results to assess the tools, we design a novel instantiating of  the general cost-benefit framework of Iannacone \& Bridges \cite{iannacone2020quantifiable} to quantifiably assess and compare the network-level tools (question Q(2)). 
This method simulates the cost to an enterprise using the technologies by integrating detection accuracy and time-to-detect statistics (all learned from experimental testing), along with estimates of attack damage costs, labor costs, and resource costs. 
We configure the cost model to estimate costs of the two network-level tools on the emulated network. Then in a separate evaluation, we estimate the additional costs for adding the network tools to complement each endpoint tool.
Our configuration of the cost model, in particular in evaluating complementary detectors used in tandem, is novel. 
It gives new vantage points for understanding the tools' efficacy  use and provides a procedure to assist future acquisition decisions. 
See Section \ref{sec:cost-model}.


\subsection{Results Summary}
\hl{The key findings from our experiment are as follows: }
\begin{itemize} 
\item These COTS detectors have nearly perfect precision but with detection rates in the 34--55\% range. 
\item Detection rates jump to $\smalltilde60\%$ for any pair of a host and a network detector; combining more than three detectors does not provide much more gain.
\item From a cost simulation analysis, the host signature-based detector is the best under our baseline assumptions. 
\item When assuming that ``hard'' malware (i.e., zero-days, polyglots, APT-style) files will incur larger damage costs than $n-$day files, ML-based detectors provide greater savings according to our cost model. 
\item Substantial savings are provided when adding one of the network-based detectors, with the dynamic detector providing the largest savings (for both host detectors).
\end{itemize}

We conclude this paper with a discussion in Section \ref{sec:did-discussion} of the overall takeaways and the limitations of this work. 
These findings are mapped to takeaways for (1) researchers wanting to strengthen the state of malware defense and (2) SOCs considering the purchase of a malware detector.

\subsection{Contributions} 
\hl{Our research is one of the few academic studies providing  empirical evaluation of commercially available ML-based malware detection tools, and we believe it is the largest academic evaluation of such tools to date---using a corpus of more than 3,500 files (in comparison to 1,000 files in prior work on COTS {\cite{fleshman2018static}}).}
\hl{More notably, it provides a unique set of files with a wide variety of file types of varying difficulty to detect, illuminating strengths and weaknesses of the tools.  It is in response to commercial evaluations} {\cite{av-test, selabs-test}} \hl{that do provide results of COTS malware detectors on larger corpora but fail to differentiate among the tools.}
\hl{Our contribution is also in our analysis of the results, considering complementary detectors and multiple applications of a cost-benefit analysis.}
We provide two novel contributions to the general cost model framework of Iannacone \& Bridges \cite{iannacone2020quantifiable}. 
We show how to use a large number of benign/malware samples to gain accuracy in detection accuracy statistics, which are needed inputs to the model, while scaling the model to represent a realistic benign/malware ratio in cost calculations. 
Our second configuration of the cost model estimates the additional cost/savings of a network-level tool, which is different and potentially more appropriate or useful than originally proposed. 
\hl{Code for our cost model configuration is provided, see footnote on first page.} 

Our statistical results provide empirical verification that contributes to a better understanding of the state of commercial malware detection in practice, a problem that has received little attention by researchers; 
in particular, we find several key weaknesses in the four tools investigated, including surprisingly low detection rates (but high precision),  a complete failure to detect polyglot files, and varying detection capabilities across file categories and, for network-level tools, across protocols. 
This helps confirm (at least our, i.e., the authors') suspicions about efficacy of commercial detectors.   
Finally, we map our findings to takeaways first for researchers, giving prioritized research directions to enhance detection in practice, and second for SOC personnel, giving guidance to consider when purchasing such tools. See the discussion in Section \ref{sec:did-discussion}. 

   \section{Related Work}
A large body of work proposes and tests ML algorithms for malware detection, and surveys have collected and organized these works. 
However, to our knowledge,  few works seek to evaluate commercial, especially, ML-based malware detection, and none provides cost-benefit analysis informed by tool-specific detection statistics learned from experimental results for reasoning about these detectors. 
In addition, the most thorough existing evaluation of malware detection tools in the literature uses a corpus of  1,000 samples, compared with our evaluation of more than 3,500 samples. 
Finally, no previous work has the efficacy of pairing complementary detectors as suggested by the ``defense-in-depth'' idea. 
Here we address commercial evaluations, which are much larger than ours, but do not provide differentiation of tools---many tools get near perfect results---inducing skepticism of the efficacy of these evaluations. 

\subsubsection{COTS Evaluations}
Early work by Christodorescu \& Jha \cite{christodorescu2004testing} presents an evaluation of three COTS malware detectors on eight malware, with a goal of studying what signatures a blackhat hacker can learn about a blackbox detector. 
Notably, these authors summarize findings by exclaiming ``From
our experimental results we conclude that the state of the art for malware detectors is dismal!''
The two studies that focus specifically on the capability of existing tools to detect malware use a corpus of 200 samples (100 malicious and 100 benign)~\cite{aslan2017investigation} and 29 samples (all malicious)~\cite{pandey2014performance}, respectively.
The more thorough of these evaluations, conducted by Aslan and Samet~\cite{aslan2017investigation}, finds that static and dynamic analyses work better in tandem than either one on its own and that, in general, dynamic analysis outperforms static. 
Neither of these works focus on the benefit that ML provides for malware detection. 

A large body of work has emerged to evaluate whether adversarial machine learning (AML)  can be applied to perturb files  to ``trick'' a detector \cite{ling2021adversarial}, although the focus is to develop the AML techniques rather than evaluate commercial detectors. 
Notably, Fleshman et al. \cite{fleshman2018static} test four commercial antivirus (AV) technologies and self-made ML classifiers on perturbations of malware. 
Similar to this study, time constraints limit Fleshmen et al. to a corpus of 1,000 malware files, and further limitations are noted: ``all the malware ... has been known for some time and likely been used by the AV companies in their product updates.''
Overall, the results exhibit that the two (not commercial) ML-based classifiers are more robust to adversarial perturbations than the four commercial detectors.

VirusTotal\footnote{\url{www.virustotal.com}} (VT) is a leading threat intelligence source that tracks malware and provides determinations of 65+ commercial detection engines  for any submitted file.  
Previous work has involved VT data on files to study commercial malware detection engines. 
Zhu et al. \cite{zhu2020measuring} track labels (benign/malicious) through VT of these 65+ vendors for more than 14K files over time and focus on the label dynamics. 
A primary finding is that commercial detection engines represented on VT are finicky, changing detection results often; nevertheless,  categorizing files based on a threshold voting schemes can be reliable for many thresholds that are intermediate in magnitudes. 
A portion of the paper uses files for which Zhu et al. know ground truth (benign/malicious) and is most related to our study. 
Specifically,  Zhu et al. create 120 zero-day ransomware samples (60 obfuscations of two known ransomware) and 236 benignware samples to test the commercial engines through VT. 
Observations include that detection rates are better over time, but for these samples, true positive \textit{and false positive} rates vary from approximately 25--100\%--which indicates wide variation in the commercial tools but also much worse precision than found in our study. 
Finally, Zhu et al. compare desktop version of 36 AV technologies against their VT counterparts, finding VT versions have on average higher recall but more false positives. 

Prior works have evaluated the efficacy of open-source network-based intrusion detection systems (IDSs), specifically Snort, Suricata, and Zeek~\cite{shah2018performance,murphy2019comparing,thongkanchorn2013evaluation}. Although these are commonly used in SOCs, they are not malware detectors.

\subsubsection{Market Reports by Companies} 
\label{sec:market-reports}
\hl{Market reports are used by security operations to learn about products in the cyber technology marketplace; although these are not peer-reviewed academic research, we include information on relevant market reporting companies to provide context.
Gartner (\hbox{\url{gartner.com/en/information-technology}}), a popular provider of market reports for the cyber technology space, curates market summaries composed of company profiles of the vendors, ratings and comments from end users of the products, information from the vendors' white papers, and often scores on a ``magic quadrant,'' proving quantification of the technologies' ``ability to execute'' and ``completeness of vision'' {\cite{gartner-mq}}. 
Gartner does not provide any scientific testing of the efficacy of detection tools. 
}

\hl{
Evaluations of IDSs are provided by commercial companies {\cite{fleshman2018static}}. 
MITRE ATT\&CK provides an evaluation online {\cite{mitre}} in which they examined 21 IDS products against the APT3 and APT29 threat groups. 
This evaluation includes custom malware and alternate execution methods. 
However, both the APT29 and APT3 include only two scenarios, for a total of four attack scenarios, so very little malware is actually executed. 
These evaluations do not specifically focus on the capability of the tools under test to detect malware but rather on their capability to detect realistic attack campaigns at any stage of execution. 
AV-TEST (\hbox{\url{av-test.org}}) and SE Labs (\hbox{{\url{selabs.uk}}}) both provide ratings based on evaluations of the malware detectors.
Although AV-TEST and SE-Labs' endpoint test results provide a rich source of statistics, they have two  main drawbacks. 
First, evaluations are seemingly too easy, diminishing trust in the tests and inhibiting the tests' capabilities to differentiate the technologies. 
As an example, in the evaluation of April 2022 {\cite{av-test}}, of the 20 detectors evaluated, 17 receive a 6/6 protection rating, with the remaining three obtaining a 5.5/6. Digging deeper, we find that the reviewers claim the industry average for detecting zero-day files is 99.8\%. 
Similarly, the quarter 1 endpoint detector test from SE Labs {\cite{selabs-test}} finds the top (approximately ten) detectors achieve near perfect (above 99\% accuracy) in most tests. 
Comparing these results with our findings---where recall of any single detector is under ~60\% across all tested malware---we are skeptical of the differentiation power of the tests provided. 
The second drawback is inherent to statistical results; making sense and reasoning logically about  many diverse measurements for each tool is difficult and has been a problem for evaluation of cyber technologies more generally {\cite{iannacone2020quantifiable}}. 
Our work seeks similar tests for these tools but provides an approach that addresses these drawbacks by providing a variety of files to (hopefully) be sufficiently difficult and to use a cost-benefit framework to reason about many diverse measurements produced by experiments.  
}

\hl{
Notably, VT does not provide comparative analytics of detection engines {\cite{virustotal}}.
}


\subsubsection{Malware Detection Method Surveys}
Myriad papers examine existing or proposed novel malware detection methods.
Because our work focuses on evaluating the efficacy of commercially available tools, we do not discuss these works in detail.
However, in this section, we provide references to several surveys that provide thorough overviews of this literature.
A number of well-cited surveys discuss malware detection methods generally~\cite{aslan2020comprehensive,idika2007survey, ye2017survey}, and others focus more narrowly on areas such as dynamic~\cite{or2019dynamic}, static~\cite{shalaginov2018machine, souri2018state}, and mobile~\cite{qamar2019mobile} malware detection.
In their recent survey of malware detection approaches, Aslan and Samet~\cite{aslan2020comprehensive} highlight several research challenges that remain open issues in the malware detection space, including the following that are particularly applicable to our work:  no detection method can detect all unknown malware, obfuscation might prevent malware from being examined by existing tools, and false positives (FPs) and false negatives (FNs) are a problem with existing approaches. 
In our research, we quantify these issues in existing tools and suggest ways they can be alleviated, such as by using multiple approaches in tandem. 

\subsubsection{Methods of Evaluation of Detectors}
We refer the reader to Iannacone \& Bridges \cite{iannacone2020quantifiable}, which provide a survey of evaluation methods for cyber defenses, in particular intrusion detectors. 
Trends other than the usual statistical metrics (e.g., recall, precision) include incorporating time of detection into IDS metrics \cite{garcia2014empirical} and cyber competition scoring frameworks \cite{red-team-article, doupe2011hit, patriciu2009guide, reed2013instrumenting, werther2011experiences, mullins2007cyber}.  
Several significant works on security cost modeling, such as the Security Attribute Evaluation Method \cite{butler2002security}, the Return on Security Investment model \cite{sonnenreich2006return, davis2005return}, and the Information Security Risk Analysis Method \cite{karabacak2005isram}, led to the general cost-benefit framework of Iannacone \& Bridges that we build upon in Section \ref{sec:cost-model}. 
\hl{ Kondakci {\cite{Kondakci_2009}}  provides a costs analysis for the spread of malware through a network using epidemic-inspired models, which is an interesting and worthwhile work but not appropriate for our setting.} 

\subsubsection{Polyglot Files}\label{rel:poly}
Polyglot files are valid as multiple file types.
For example, a file could be a valid joint photographic group (JPG) and a valid Java archive (JAR) file, or it could be a valid portable document format (PDF) and a valid hypertext preprocessor (PHP) file.
Malware detection methods often require the file type to be known, so a polyglot file is used to prevent the correct file type association to bypass the detection mechanism.
A large body of research exists related to exploiting systems using polyglot files~\cite{bratus2016fillory, bratus2017exploitation,albertini2015abusing,magazinius2013polyglots,wolf2010omg}, as well as online tutorials on creating polyglot files\footnote{\url{https://medium.com/swlh/polyglot-files-a-hackers-best-friend-850bf812dd8a}}. 
However, there is little research focused specifically on helping IDS systems handle polyglot files.
The majority of existing research is focused on the challenges of identifying malicious PDF files~\cite{nissim2015detection, carmony2016extract}.
Although solutions such as binwalk\footnote{\url{https://github.com/ReFirmLabs/binwalk}} can be used to help identify polyglot files, it is unclear why they have not been integrated into commercial tools. 
One possible reason is that using a tool such as binwalk would have a negative impact on throughput.

\section{Experimental Design}
\label{sec:experiment}
To execute the experiments, we leveraged the NCR, an air-gapped research test bed equipped with state-of-the-art network and user emulation capabilities to facilitate the experiment \cite{ferguson2014ncr}. 
A variety of services were emulated, including email, website servers, and domain controllers, and they were used by emulated users both internal to the network and in the emulated worldwide web (external).  
\hl{We provide a network diagram in the Appendix (Figure} \ref{fig:network-diagram}). 

\hl{Traffic generation was provided on all client endpoint virtual machines (VMs) by the NCR Mantra soft
ware, a proprietary emulator built on Lariat {\cite{lariat}}. 
All clients had broadly similar traffic generation profiles and were configured to automatically and randomly browse websites (both internal and external), send and reply to emails, generate Microsoft (MS) Office documents, transfer files to and from file shares and Sharepoint sites, and run a preset list of other benign executables. All clients were set to a fixed diurnal schedule of increased daytime activity between the hours of 0700 and 1700 local time; activity outside of these hours was decreased by a factor of 10 (i.e., average inter-arrival time of traffic generation events was set to be 10 times as high).}

\hl{System time was synchronized between all devices except the network static appliance via NTP to the central NCR testbed time source. Every node with an in-band connection synchronized to an NTP server in the simulated Internet environment, which in turn connected to the central time source via the testbed control network; everything with only an out-of-band connection connected directly via the
control network to the same source. The network static appliance was supplied as a vendor-loaned server without administrative (root) access, which is required to change time settings; the offset of this appliance’s time vs. actual time was noted for compensation in after-action data analysis.
}

This environment enabled delivery of malware over the network to endpoints running the two host-level tools, so the network-level tools could (attempt to) reconstruct files from the packet stream and provide detection.  
In particular,  emulated internal and external websites, mail servers, and hosts were used to deliver files (both benign and malicious) to  internal clients through web download, email attachments, and direct connections, respectively.

\hl{The authors met with vendors of the technologies and configured the tools in advance of the NCR to ensure proper configuration. 
Configuration guides were used to duplicate the process at NCR, and, as a quality assurance check, detection results on a set of test files were confirmed to be identical before the experiment (with tools configured by the vendors) to  after setup in the NCR.} 
All technologies operated on-premises, without connections outside the internal portion of the emulated network. 
Although both host-level tools provide pre-execution conviction and quarantine, for the purpose of the experiment all tools were set to alert only (not block/prevent) to permit complementary (\texttt{OR}) 
analysis of multiple tools. 
All machines were time synchronized. 
For the network-level tools, a secure sockets layer (SSL) 
``break-and-inspect'' decryption technology was \textit{not} added to this network. Files were delivered on unencrypted protocols as the goal was to test the detectors' capabilities to detect malware without dependency on the capabilities of the decryption technology. 
(In practice, under this configuration malware sent encrypted would almost certainly be opaque to these detectors. How such a decryption technology affects the network-level detector is out of scope for this work.) 


For this experiment we evaluated the endpoint detection on both Windows 7 and 10 for an endpoint (nonserver) host, although Linux hosts and servers were present. 
Windows 7 and  10 VMs were given 4 GB of RAM and 1 CPU core, along with a large disk drive for the installed operating system. 
RHEL6 VMs were given 1 GB of RAM and 1 CPU core, also with a large disk drive for storing the operating system and associated files.
These values reflect the amount of minimum memory needed to create a VM for the specific operating system, with both the Windows VMs requiring at least 4 GB and RHEL requiring only 1 GB. 
The test bed was controlled via orchestration infrastructure leveraging virtualization software and custom wrapping software to coordinate with the VMs and manage the VM life cycle. 

    The detection technologies were each exposed to 3,536 total files (2,554/982 = 72/28\% malicious/benign) of varying categories, including 32- and 64-bit  portable executables (PE32, PE64), MS  Office, PDF, JAR, and three unusual categories: APT (described subsequently), zero-day PE (Windows PEs believed to not be available publicly), and polyglot  (files that are valid as multiple file types, e.g., ZIP and PDF). 
Most malicious files were obtained through VirusShare\footnote{\url{https://virusshare.com/}}. 
Our polyglot categories should also be never-before-seen files, and at least one of the valid file types has malicious functionality; 
\hl{specifically, we combined malicious Java JAR files with GIF and JPEG files. 
Notably, one can leverage open-source software for polyglot creation} (\url{https://github.com/corkami/mitra}). 
The zero-day files were manufactured samples that never touched infrastructure with access to the Internet. 

The goal of the APT class was to create artifacts that exemplified common APT evasion tactics. This included using signed vs. unsigned executables, cloning the PE headers of common benign executables, and advanced in-memory characteristics. 
Cobalt Strike\footnote{\url{www.cobaltstrike.com}} provided a convenient way to test these features. 
Because Cobalt Strike is a known tool, we needed to minimize the risk that our artifacts would be discovered by generic Cobalt Strike signatures. 
To this end, we built a custom obfuscator for the samples in the APT class using the Cobalt Strike artifact kit. 
The artifact kit is effectively an API that allows users to change the way Cobalt Strike generates binary artifacts, allowing advanced users to more successfully evade AV technologies. 
As a quality assurance step, we built a named pipe bypass using the artifact kit and ensured that default samples (i.e., samples without any of the APT-like evasions) generated with this obfuscation were not detected by a fifth, commonly used malware detector. 
With this obfuscation in hand, the APT class was created by combining that obfuscation with the various APT-like evasions available in Cobalt Strike. 
For those APT samples that were correctly detected in our experiment, we cannot know if the correct detection was from a generic signature (that our obfuscator failed to evade) or if indeed the APT-like evasion tactic was correctly identified behaviorally. 
\begin{wrapfigure}[12]{r}{.5\textwidth}
    \centering
    \includegraphics[width=.5\textwidth]{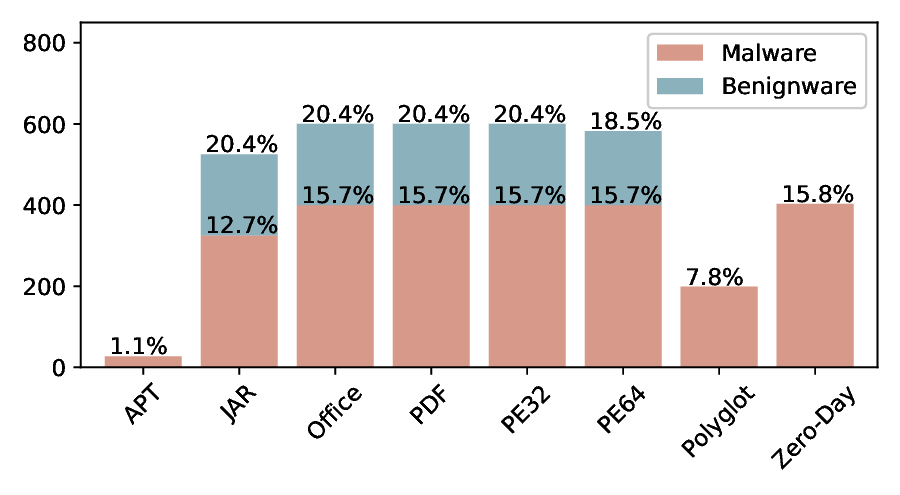}
    \caption{File category percentages}
    \label{fig:tc2-filetype-pie}
\end{wrapfigure}
\subsection{File Samples}

Our experiment used an imbalanced class, that is, a smaller number of benign files than malicious. 
As the experiment involved instantiating an entire analysis environment and waiting up to two minutes for a detection decision on each file, time limitations forced our decision to limit the benign files. 
We biased the benign set to include a set of ``tricky'' benign samples generated and donated by researchers who developed the Hyperion tool \cite{prowell2016automatic, linger2013computing}. 
These benign files are mostly C/C++ compiled utilities and were often (incorrectly)  classified as malicious in Hyperion research. 
The rest of the benign files are standard office type documents, JAR files, RPMs, DEBs, and EXEs/MSIs.

Overall, inclusion of nonpublic malware and benignware allows testing of the ML generalizability hypothesis---that it can correctly identify never-before-seen files, while the variety of malware allows examination of recall across file types and categories.


\begin{table*}[ht]
\caption{Individual sensor results itemized. \hl{Categories are disjoint in this table: all 403 zero-day files are PE32 files but are \textit{not} counted in the \textbf{PE32} row; similarly, all polyglot files are malicious JARs but are \textit{not} counted in the \textbf{JAR} row. The \textbf{Total} row provides the statistics computed on all files. }}
\begin{adjustwidth}{-2in}{-2in}
\centering
\label{tab:tc2:comparative-analysis}
\begin{tabular}{lllllllllllll}
\toprule
\textbf{Category} & \multicolumn{3}{c}{\textbf{Host Signature}} & \multicolumn{3}{c}{\textbf{Host ML}} & \multicolumn{3}{c}{\textbf{Network Static}} & \multicolumn{3}{c}{\textbf{Network Dynam.}} \\
(\#mal./\#ben.) &                 Prec. & Recall &    F1 &          Prec. & Recall &   F1 &           Prec. & Recall &   F1 &            Prec. & Recall &   F1 \\
\cmidrule(lr){2-4}\cmidrule(lr){5-7}\cmidrule(lr){8-10}\cmidrule(lr){11-13}
\textbf{APT} (27/0)       &                     1 &  0.667 &    0.8 &              1 &   0.63 & 0.773 &               1 &  0.296 & 0.457 &                1 &  0.556 & 0.714 \\
\rowcolor[HTML]{EFEFEF}
\textbf{JAR} (325/200)    &                     1 & 0.0892 &  0.164 &              0 &      0 &     0 &               1 &  0.249 & 0.399 &                1 &  0.209 & 0.346 \\
\textbf{Office} (400/200) &                     1 &  0.782 &  0.878 &              0 &      0 &     0 &               1 &  0.873 & 0.932 &                1 &  0.735 & 0.847 \\
\rowcolor[HTML]{EFEFEF}
\textbf{PDF} (400/200)    &                     1 &  0.512 &  0.678 &              0 &      0 &     0 &               1 &   0.51 & 0.675 &            0.996 &  0.565 & 0.721 \\
\textbf{PE32} (400/200)   &                     1 &  0.505 &  0.671 &          0.922 &  0.855 & 0.887 &           0.814 &   0.69 & 0.747 &            0.977 &  0.642 & 0.775 \\
\rowcolor[HTML]{EFEFEF}
\textbf{PE64} (400/182)   &                     1 &  0.233 &  0.377 &              1 &  0.777 & 0.875 &               1 &  0.833 & 0.909 &                1 &  0.757 & 0.862 \\
\textbf{Polyglot} (199/0) &                     0 &      0 &      0 &              0 &      0 &     0 &               0 &      0 &     0 &                0 &      0 &     0 \\
\rowcolor[HTML]{EFEFEF}
\textbf{Zero-Day} (403/0) &                     1 & 0.0347 & 0.0671 &              1 &  0.486 & 0.654 &               1 &  0.395 & 0.566 &                1 &  0.553 & 0.712 \\
\cmidrule(lr){2-4}\cmidrule(lr){5-7}\cmidrule(lr){8-10}\cmidrule(lr){11-13}
\textbf{Total} (2554/982) &                     \textbf{1} &  0.342 &   0.51 &          0.968 &  0.339 & 0.502 &           0.957 &  \textbf{0.552} &   0.7 &            0.995 &  0.543 & \textbf{0.702} \\
\bottomrule
\end{tabular}
\end{adjustwidth}
\end{table*}

\subsection{Tested Technologies} 
\label{sec:tools}
All technologies tested are established malware detection products from well-known companies in the cyber tech market place. 
The three ML-based tools used in the test are of type  host ML, network static, and network dynamic detectors. 
\hl{All ML technologies came pretrained.} 
A popular endpoint signature-based detector was used as the baseline. 
The endpoint ML detector claims to use no signatures; therefore, it allow us to compare signature-only to ML-only detection at at the host level. 
Two network-level tools provide ML-guided detection, but one is static and the other is dynamic.
To test their technologies, we agreed to keep the  vendors and their intellectual property confidential.

\textbf{Host-Level, Signature-Based Detector:}
This proprietary detection tool uses static code analysis to compare hashes and portions of the file's binary with signatures, known malicious/benign hashes, and code snippets. 
This standard approach for an AV tool promises high precision and fast detection but an inability to identify novel, even slightly changed malicious files. 
In general, porting of signatures to each host requires relatively large host memory usage but facilitates fast and computationally inexpensive detection. 

\textbf{Host-Level, ML Detector:}
This host-based AV tool promises pre-execution file conviction for certain file types. 
The binary classification uses supervised ML trained on large quantities of both malicious and benign files. 
According to vendor reports, the product requires 150 MB of memory on each host and claims that detection incurs computational expense similar to that incurred by taking a hash. 

\textbf{Network-Level, Static, ML Detector:} 
The network-level, static detection tool was designed to  passively detect (not block) both existing and new/polymorphic attacks before a breach, on the wire, in near real time, with an on-premises solution.  
This technology centers on a binary (i.e., benign/malicious) classification of files and code snippets extracted from network traffic. 
All features of the reconstructed files and code snippets result from static analysis. 
This product's sensors sit at a tap between the firewall and switch and between the router and hosts and rely on a variety of open-source tools for both feature extraction and classification; that is, in addition to their proprietary ML classifier, open-source file conviction technologies are used. 

\textbf{Network-Level, Dynamic, ML Detector:} 
The network-level dynamic detector reconstructs files from the network data stream and, 
using a central analytics server that resides either on-premises or in the cloud, builds both static and dynamic features by running the file in sandboxes and emulation environments for analysis. 
The vendor claims the tool accommodates many file types including executable, DLL, Mach-o, Dmg, PDF, MS Office, Flash, ISO, ELF, RTF, APK, Silverlight, Archive, and JAR. 
Features are then fed to a binary supervised classifier. 
Presumably, network-level and behavioral features are included as well. 
Detection timing was quoted as up to 20 s to a few minutes, which, based on vendor reports, is much (i.e., orders of magnitude) slower  than competitors, although most competitors use static analysis only. 
This highlights the fact that, as with the network-level static detector tested, this product is primarily for detection, not  prevention.

\section{Statistical Results}
\label{sec:stats-results}

Here, we provide the usual accuracy  and conviction time statistics in aggregate, broken down by file type, and, for the network-level tools, by  delivery protocol. 
Furthermore, we present results for each detector side by side and in complementary combinations.  

\subsection{Individual Sensor Results} 
Table \ref{tab:tc2:comparative-analysis} itemizes the results of each malware analysis technology, with rows per file type except the bottom row, which gives overall detection statistics. 
Because a focus of this experiment is to identify the ways in which ML-based technologies add value, we consider the host signature-based sensor as a baseline.  
The host signature-based tool demonstrates perfect precision but poor recall, pulled down  by its inability to detect never-before-seen malware (0\% polyglots, 3.5\% zero-days detected). 
The tool performs relatively well with respect to MS Office files and APT files and fairly well with respect to executable and PDF files.  
Good performance on executable, MS Office, and PDF file types is not surprising because these are the most common packaging for malware and, therefore, have the broadest set of signatures.  
As expected, the host signature-based tool performed poorly on JAR files and zero-day samples, where there are fewer relevant signatures, or in the case of zero-day samples, no existing signatures.

The good performance on the APT category was a surprise.  
It is likely that the Cobalt Strike 
framework used for packaging the APT malware samples just happens to be well known to the signature set used in this host signature-based sensor. 
The host signature-based tool's perfect precision simplifies this analysis, as recall becomes the primary metric of comparison.
Performing worse in overall recall, the host ML tool has 0\% detection for all non-PE files. 
The two network-level tools perform quite similarly, with good ($>$95\%) precision and both with better (but still alarmingly poor) recall of 55\%, giving F1 near 70\%. 

The network static tool 
achieves a recall of $\smalltilde$20\% greater than the host-based tools. 
As this tool uses a suite of ML- and signature-based subclassifiers, these results suggest that complementary signature and ML tools might work well.   
This increase in recall is shadowed by a higher quantity of FP alerts.   
Nearly identical overall statistics are achieved by the lone dynamic tool. 
\hl{Our results imply that network-level tools  do  increase recall substantially with comparable precision to the host ML tool and only slightly worse precision to the host static tool.}

Overall, regarding Table \ref{tab:tc2:comparative-analysis}, there is no clear ``winner.''
Precision values are high for all four tools, meaning all sensors performed with relatively few false alerts.  
Perhaps the biggest takeaway of these results is the utter failure of these tools to detect most malware. 
Recall of 33--55\% is both disappointing and alarming, with roughly 37\% of the malware used in the exercise undetected. Furthermore, recall varies wildly across file categories for each tool. 
Notably, most operations use a sole endpoint detector, and both of these state-of-the-art commercial detectors identified only a third of the malware used in this test. 
Polyglots are a glaring problem for these four tools, as no such file was detected. 
This suggests that the approach to detection is dependent on determination of file type. 

Considering the zero-day category, we see that all three ML-based tools (host ML and both network tools) increase recall by more than a factor of ten. 
This quantifies our first hypothesis, at least for these tools, that ML, as incorporated into modern detection products, can help generalize to never-before-seen files. 
Simply put, in our results, the detection rates increase by approximately ten times but at the cost of precision---specifically precision on PE32 files---which drops by approximately 10\%.  
Considering only the zero-day PEs and the APT files, the network dynamic tool is (unsurprisingly) superior. 
Comparing this result with the amount of time needed to detect will put these figures in context.

\begin{figure}[t]
    \centering
    \includegraphics[width=.495\textwidth]{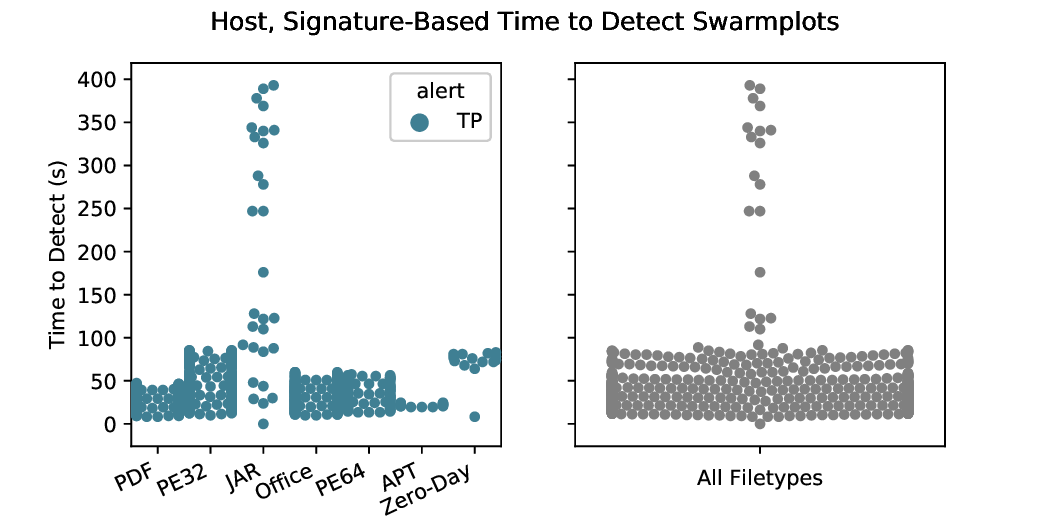}
    \includegraphics[width=.495\textwidth]{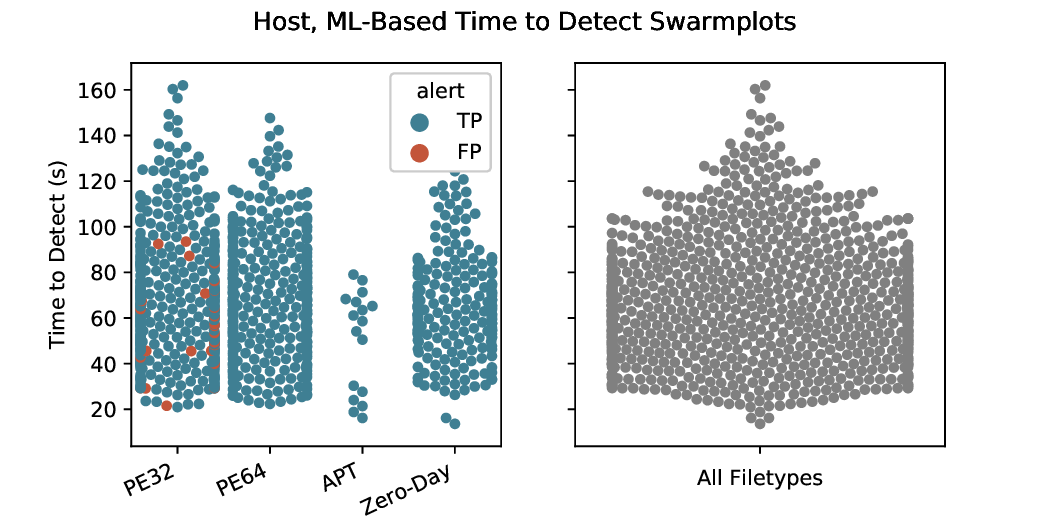}
    \includegraphics[width=.495\textwidth]{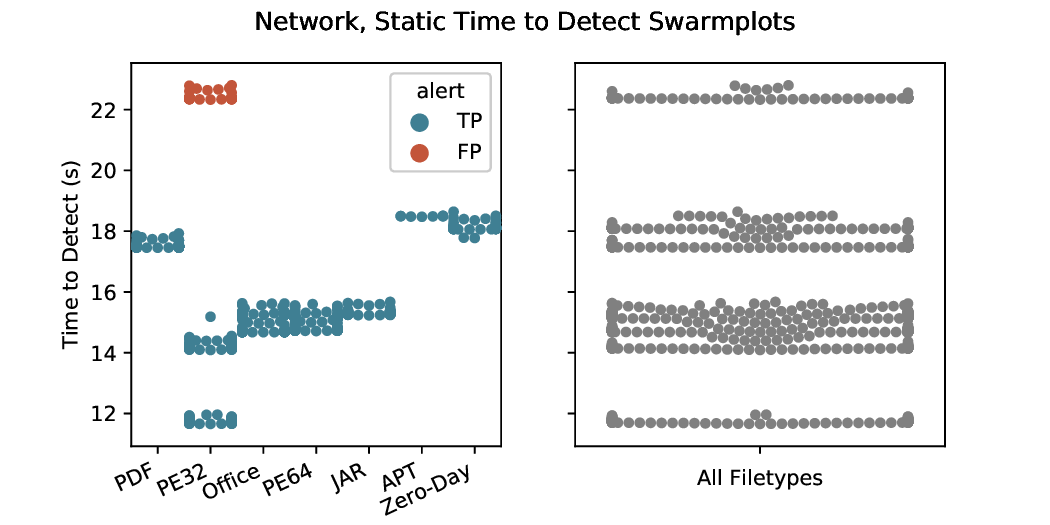}
    \includegraphics[width=.495\textwidth]{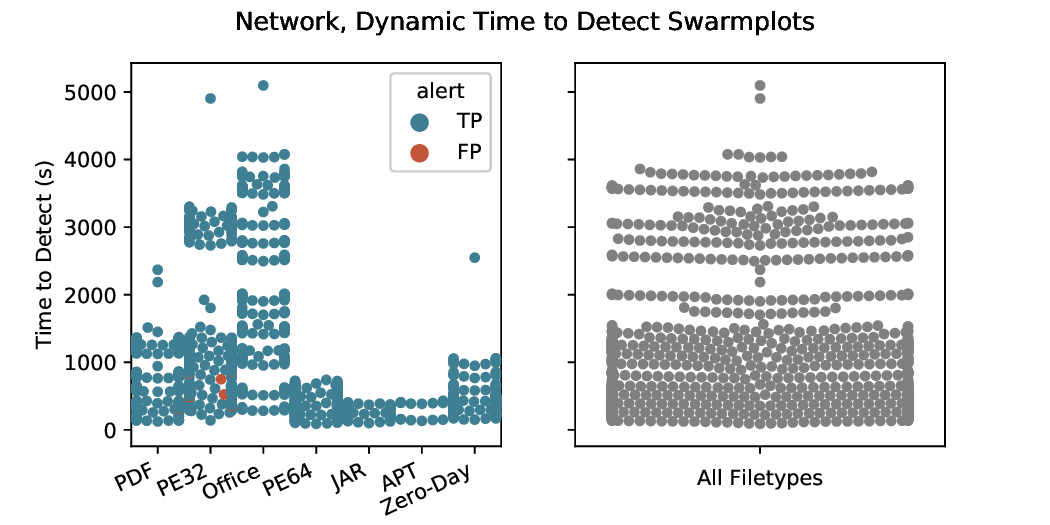}
    \caption{Swarm plots show time to detection for each file type. Left plots show per-file-category results with blue/orange data for true/false alerts. Right plots show all file categories and depict the time-to-detection distribution.  
    \textbf{Top:} (Left) As the signature-based detector  has perfect precision, there are no FPs (no orange data points). (Right) Host ML detector does not detect many file categories and has many FPs in PE32 files. 
    \textbf{Bottom:} (Left) Network static detector has nearly identical time to detection for each file type except PE 32, the only type with false alerts, and all have relatively long analytic time. 
    (Right) The network dynamic detector has very few false alerts, but time to detect is one to two orders of magnitude longer. 
    }
    \label{fig:tc2:malware-analysis-times}
\end{figure}

\subsubsection{Detection Latency}
\label{sec:tc2:analysis-latency}

In addition to accuracy, timely detection is imperative to effective defense against malware.  
Figure \ref{fig:tc2:malware-analysis-times} shows each tool's time from file delivery until alert 
both per file type with true and false positives (TPs/FPs) indicated by color and in aggregate across all files to visualize the overall alert time empirical distribution. \hl{Detection latency is calculated from the timestamp generated when the file transmission is completed until the timestamp of a given tool's alert on the respective file and encompasses all processing, including any file carving or decompression required.}
Tools' average conviction times rank as expected, but the results hold some unexpected takeaways. 
The network static tool has the fastest average conviction time of 15.3s, perhaps due to the resources available for a network appliance (much more than an application on a host) and static technique (generally faster than dynamic detection). 
Interestingly, all its FPs occur in PE32 files that take maximum times to alert for this tool. (Without visibility into the detection mechanism of the tool, we cannot diagnose this phenomenon. We notified the vendors of this anomaly.)
Both host-based tools take at most a couple minutes to alert, with the ML-based tool slower on average. 
Anomalously, the signature-based tool JAR files exhibit wide variance in time to alert. 
Recall that the host ML detector does not alert on many file categories and has many FPs on PE32 files, whereas the host signature-based tool has no FPs. 
The network dynamic tool has a \textit{much} greater average conviction time of 1,039 s. 
This is understandable given the process of creating a sandbox environment and then executing the file in that environment to make a conviction decision.

We dug into the detection latency of these network-level tools in an additional experiment. 
Both vendors provided updated network appliances, and we tested their time-to-detect capabilities in a separate, more realistic experiment. 
Each tool received a mirrored feed of a real network's traffic blended with emulated network traffic, allowing more than 2,000 malware samples to be delivered safely to VMs but spread over a few hours. 
While seeing data rates of up to 7 GB/s, the tools achieved median detection times, 
\hl{ from file delivery until alert,} 
of under 1 s for the network static tool and 258 s = 4.3 m for the network dynamic tool. (Median is reported here because of outliers in the statistics that affected the mean.) 
The network-level tools had a significant bottleneck in the computational overhead of decompressing files. 
Introducing emulated network data had a positive effect on detection latency as it significantly decreased the rate at which compressed files were delivered to the tools despite dramatically increasing the amount of raw data to process---the tools received more network traffic but a decreased rate of compressed files to process.
These results show that, indeed, there is a latency cost for using at least this dynamic analysis tool and that both the network tools' latencies can increase by an order of magnitude if files cross the network too rapidly relative to the server's computational resources.

\subsubsection{Network-Level Detectors Efficacy Across Protocols}
\label{sec:protocols}
\begin{wraptable}[]{r}{.33\textwidth}
    \centering
    \caption{Detection counts for network-level detectors on 200 malware samples each sent on five protocols.}
    \label{tab:detection-across-protocols}
    \begin{tabular}{lcc}
        \toprule
          &  \textbf{Dynamic} &  \textbf{Static} \\
        \midrule
              FTP &      0/200 &        9/200 \\
             \rowcolor[HTML]{EFEFEF} HTTP &    145/200 &      156/200 \\
              SMB &      0/200 &       38/200 \\
             \rowcolor[HTML]{EFEFEF} SMTP &    180/200 &        N/A \\
              TCP &      0/200 &        0/200 \\
        \bottomrule
    \end{tabular}
\end{wraptable}
The two network-level tools rely on extracting files from traffic 
\hl{and claimed accommodation of multiple protocols. 
As such, we ran a small test using multiple protocols.} 
Our results revealed that visibility  into those network streams was not as reliable as expected. 
All of the 3,536 file samples used in the analysis were delivered over the HTTP protocol. 
To test the network tools' capabilities across protocols,  200 of the malware samples were also delivered over four additional protocols. \hl{To deliver files over multiple protocols, web downloads (HTTP), external file transfers (FTP), internal file transfer (SMB),  attachments to  emails (SMTP), and clear-text TCP via Netcat were used.  
Netcat was used to test a vendor's claim that it could detect files sent in this manner.} 
We present the counts of correctly detected samples for these 200 malware samples on the five total protocols in Table \ref{tab:detection-across-protocols}. 
The network static tool has an email client that was not used in this analysis, and so its performance on email was not determined. 
All network transmissions were unencrypted. 
Both network IDS tools require de-/un-encrypted traffic. 
These results show that the tools' capabilities to detect varies greatly by protocol, and this does not include the complexity induced by the realistic assumption that  these tools would sit downstream of a companion decrypter.

\begin{table*}[t]
\caption{\hl{Top half of table: itemized complementary malware detection results }  ($result(A \texttt{ OR } B)$). \hl{Bottom half of table: Difference of the first (lone, base, host) detector's  results in Table {\ref{tab:tc2:comparative-analysis}} from the paired complementary detectors' results  in the top half of this table}  ($result(A \texttt{ OR }  B) - results(A)$).} 
\begin{adjustwidth}{-2in}{-2in}
\centering
\label{tab:tc2:compl-analysis}
\begin{tabular}{llllllllllllllll}
\toprule
\textbf{Category } & \multicolumn{3}{c}{\textbf{HostSig$\cup$HostML}} & \multicolumn{3}{c}{\textbf{HostSig$\cup$NetStat}} & \multicolumn{3}{c}{\textbf{HostSig$\cup$NetDyn}} & \multicolumn{3}{c}{\textbf{HostML$\cup$NetStat}} & \multicolumn{3}{c}{\textbf{HostML$\cup$NetDyn}} \\
(\#mal/\#benign)  &  Prec. & Recall &    F1 &   Prec. & Recall &    F1 &   Prec. & Recall &    F1 &   Prec. & Recall &    F1 &   Prec. & Recall &    F1 \\
\cmidrule(lr){2-4}\cmidrule(lr){5-7}\cmidrule(lr){8-10}\cmidrule(lr){11-13}\cmidrule(lr){14-16}
\textbf{APT} (27/0)       &                        1 &   0.67 &  0.8 &                            1 &   0.67 &  0.8 &                            1 &   0.67 &  0.8 &                          1 &   0.63 & 0.77 &                          1 &   0.63 & 0.77 \\
\rowcolor[HTML]{EFEFEF}
\textbf{JAR} (325/200)    &                        1 &  0.089 & 0.16 &                            1 &    0.3 & 0.46 &                            1 &   0.25 &  0.4 &                          1 &   0.25 &  0.4 &                          1 &   0.21 & 0.35 \\
\textbf{Office} (400/200) &                        1 &   0.78 & 0.88 &                            1 &   0.93 & 0.96 &                            1 &   0.91 & 0.95 &                          1 &   0.87 & 0.93 &                          1 &   0.73 & 0.85 \\
\rowcolor[HTML]{EFEFEF}
\textbf{PDF} (400/200)    &                        1 &   0.51 & 0.68 &                            1 &   0.57 & 0.73 &                            1 &   0.58 & 0.73 &                          1 &   0.51 & 0.68 &                          1 &   0.56 & 0.72 \\
\textbf{PE32} (400/200)   &                     0.93 &   0.91 & 0.92 &                         0.85 &   0.86 & 0.86 &                         0.98 &   0.83 &  0.9 &                       0.85 &   0.93 & 0.89 &                       0.92 &    0.9 & 0.91 \\
\rowcolor[HTML]{EFEFEF}
\textbf{PE64} (400/182)   &                        1 &   0.82 &  0.9 &                            1 &   0.86 & 0.92 &                            1 &    0.8 & 0.89 &                          1 &   0.86 & 0.93 &                          1 &   0.81 &  0.9 \\
\textbf{Polyglot} (199/0) &                        0 &      0 &    0 &                            0 &      0 &    0 &                            0 &      0 &    0 &                          0 &      0 &    0 &                          0 &      0 &    0 \\
\rowcolor[HTML]{EFEFEF}
\textbf{Zero-Day} (403/0) &                        1 &   0.49 & 0.65 &                            1 &    0.4 & 0.57 &                            1 &   0.56 & 0.72 &                          1 &    0.5 & 0.67 &                          1 &   0.62 & 0.77 \\
\cmidrule(lr){2-4}\cmidrule(lr){5-7}\cmidrule(lr){8-10}\cmidrule(lr){11-13}\cmidrule(lr){14-16}
\textbf{Total} (2554/982) &                     0.98 &   0.57 & 0.72 &                         0.96 &   \textbf{0.61} & 0.75 &                            \textbf{.996} &   \textbf{0.61} &\textbf{ 0.76} &                       0.96 &   \textbf{0.61} & 0.75 &                       0.98 &    0.6 & 0.75 \\



\cmidrule(lr){2-16}
\cmidrule(lr){2-16}
& \multicolumn{3}{c}{\textbf{HostSig$\cup$HostML}} & \multicolumn{3}{c}{\textbf{HostSig$\cup$NetStat}} & \multicolumn{3}{c}{\textbf{HostSig$\cup$NetDyn}} & \multicolumn{3}{c}{\textbf{HostML$\cup$NetStat}} & \multicolumn{3}{c}{\textbf{HostML$\cup$NetDyn}} \\
 & \multicolumn{3}{c}{\textbf{ $\Delta$HostSig}} & \multicolumn{3}{c}{\textbf{ $\Delta$HostSig}} & \multicolumn{3}{c}{\textbf{ $\Delta$HostSig}} & \multicolumn{3}{c}{\textbf{ $\Delta$HostML}} & \multicolumn{3}{c}{\textbf{ $\Delta$HostML}} \\
(\#mal/\#benign) &                                    Prec. & Recall &   F1 &                                     Prec. & Recall &    F1 &                                    Prec. & Recall &    F1 &                                      Prec. & Recall &     F1 &                                   Prec. & Recall &    F1 \\
\cmidrule(lr){2-4}\cmidrule(lr){5-7}\cmidrule(lr){8-10}\cmidrule(lr){11-13}\cmidrule(lr){14-16}
\textbf{APT} (27/0)       &                                        0 &      0 &    0 &                                         0 &      0 &     0 &                                        0 &      0 &     0 &                                          0 &      0 &      0 &                                       0 &      0 &     0 \\
\rowcolor[HTML]{EFEFEF}
\textbf{JAR} (325/200)    &                                        0 &      0 &    0 &                                         0 &   0.21 &  0.29 &                                        0 &   0.16 &  0.24 &                                          1 &   0.25 &    0.4 &                                       1 &   0.21 &  0.35 \\
\textbf{Office} (400/200) &                                        0 &      0 &    0 &                                         0 &   0.15 & 0.08 &                                        0 &   0.12 & 0.07 &                                          1 &   0.87 &   0.93 &                                       1 &   0.73 &  0.85 \\
\rowcolor[HTML]{EFEFEF}
\textbf{PDF} (400/200)    &                                        0 &      0 &    0 &                                         0 &  0.06 & 0.05 &                                  0 &  0.07 & 0.06 &                                          1 &   0.51 &   0.68 &                                       1 &   0.56 &  0.72 \\
\textbf{PE32} (400/200)   &                                   -0.07 &    0.4 & 0.24 &                                     -0.15 &   0.36 &  0.18 &                                   -0.02 &   0.32 &  0.23 &                                     -0.07 &   0.07 & 0 &                                 0 &  0.05 & 0.02 \\
\rowcolor[HTML]{EFEFEF}
\textbf{PE64} (400/182)   &                                        0 &   0.59 & 0.53 &                                         0 &   0.63 &  0.55 &                                        0 &   0.56 &  0.51 &                                          0 &  0.09 &  0.05 &                                       0 &  0.04 & 0.02 \\
\textbf{Polyglot} (199/0) &                                        0 &      0 &    0 &                                         0 &      0 &     0 &                                        0 &      0 &     0 &                                          0 &      0 &      0 &                                       0 &      0 &     0 \\
\rowcolor[HTML]{EFEFEF}
\textbf{Zero-Day} (403/0) &                                        0 &   0.45 & 0.59 &                                         0 &   0.36 &   0.5 &                                        0 &   0.52 &  0.65 &                                          0 &  0.02 &  0.01 &                                       0 &   0.14 &  0.11 \\
\cmidrule(lr){2-4}\cmidrule(lr){5-7}\cmidrule(lr){8-10}\cmidrule(lr){11-13}\cmidrule(lr){14-16}
\textbf{Total} (2554/982) &                                    -0.02 &   0.23 & 0.21 &                                    -0.04 &   0.27 &  0.24 &                                  0 &   0.27 &  0.25 &                                     0 &   0.27 &   0.25 &                                   0.01 &   0.27 &  0.25 \\
\bottomrule
\end{tabular}
\end{adjustwidth}
\end{table*}

\subsection{Complementary Detection}
\label{sec:tc2:heterosensinganalysis}

In light of the surprisingly low detection rates (33--55\%, see Table \ref{tab:tc2:comparative-analysis}), we consider simulated results of using pairs of tools together by taking the logical \texttt{OR} of binary classifiers. 
Tool pairs will score well if they alert on different malware---thereby increasing recall---but overlap on false alerts. 
Specifically, we consider pairing the two host-based detectors together to see if host signature and ML techniques are a good combination. 
Then, we consider combining each host tool paired independently with each network-level tool. See Table \ref{tab:tc2:compl-analysis}. 




Complementing the host signature-based tool with other technologies increases recall dramatically for PE32, PE64, and zero-days files. 
Precision of the host signature tool is minimally sacrificed only for PE32 files in these pairings. 
Similarly, as the host-ML tool does not alert on many file categories,  substantial gains in detection capabilities are possible by adding any other tool.
As a final baseline, we considered the union of all four tools (Table \ref{tab:all-four-unioned}), which only slightly increases recall with little change to precision over these pairs. 
In short, combinations of either the host-level tool with one other malware detector yields much better recall with little impact to precision, and only pairs of tools are needed---unioning three or more detectors provides little gain.
\begin{wraptable}[15]{r}{.4\textwidth}
\caption{\hl{
Results of complementary detection (logical union) of all four detectors.
}}
\centering
\label{tab:all-four-unioned}
\begin{tabular}{llll}
\toprule
\textbf{Category} & \multicolumn{3}{c}{\textbf{All Four Detectors}} \\
(\#mal/\#benign)  &          Prec. & Recall &   F1 \\
\cmidrule(lr){2-4}
\textbf{APT} (27/0)       &              1 &   0.67 &  0.8 \\
\rowcolor[HTML]{EFEFEF}
\textbf{JAR} (325/200)    &              1 &    0.3 & 0.46 \\
\textbf{Office} (400/200) &              1 &   0.93 & 0.96 \\
\rowcolor[HTML]{EFEFEF}
\textbf{PDF} (400/200)    &              1 &   0.58 & 0.73 \\
\textbf{PE32} (400/200)   &           0.85 &   0.94 & 0.89 \\
\rowcolor[HTML]{EFEFEF}
\textbf{PE64} (400/182)   &              1 &   0.89 & 0.94 \\
\textbf{Polyglot} (199/0) &              0 &      0 &    0 \\
\rowcolor[HTML]{EFEFEF}
\textbf{Zero-Day} (403/0) &              1 &   0.63 & 0.77 \\
\cmidrule(lr){2-4}
\textbf{Total} (2554/982) &           0.96 &   0.67 & 0.79 \\
\bottomrule
\end{tabular}
\end{wraptable}

\section{Cost Model Evaluation}
\label{sec:cost-model}
Although the aforementioned metrics provide a valuable statistical summary of these sensors' performance, recent work by Iannacone \& Bridges \cite{iannacone2020quantifiable} provides a cost--benefit framework that models the real-world implications of using a particular tool, yielding a single, comparable cost metric to more easily reason out the tradeoffs implied in Section \ref{sec:stats-results}.
For example, the network dynamic tool has an average detection time of 1,039.0 s, an FP rate of 0.71\%, and a TP rate of 54.3\%, whereas the network static tool detects on average in 15.3 s (faster) with an FP rate of 6.42\% (worse) and TP rate of 55.21\% (about the same). 
It is difficult to know which tool is best for a given SOC to adopt based on these basic statistics. 
It is even more difficult to choose the best pair of tools.

To assist with these determinations, we configured the cost 
model to evaluate these tools as standalone detectors (Section \ref{sec:result-one-tool}). This follows the original cost-model work \cite{iannacone2020quantifiable} but makes one novel contribution to the basic model. Since the cost model as originally proposed depends on the accuracy statistics of the tool, we test with a relatively large number of both malicious and benign files to accurately estimate these inputs.
Our change to the model is to compute average costs per benign-/malware, then scale the costs to respect two new model inputs, namely the expected number of files per year times the ratio of benign-/malware expected in the wild. 
This allows us to gain accuracy in both the detection  statistics and the cost model. 
In Section \ref{sec:savings-addition} we  create a new version of the model to compute savings of adding a network-based malware detector, assuming a host-based detector is in place. 
This new estimate is created to answer (Q2). 
\begin{wraptable}[12]{r}{.33\textwidth}
    \centering
   \centering
   \caption{Initial (one-time) server or appliance and ongoing (annual) subscription costs.}
    \begin{threeparttable}
    \begin{tabular}{lcc}
    \toprule
    & \textbf{Initial} & \textbf{Ongoing}\\
    \midrule
     Host Sig.  & \$2K  & \$8K/y \\ 
    \rowcolor[HTML]{EFEFEF}  Host ML  & \$6.5K & \$35K/y\\
     Net. Static & \$15K & \$20K/y\\
    \rowcolor[HTML]{EFEFEF}  Net. Dynam.  & \$23K & \$16K/y\\
    \bottomrule
    \end{tabular}
    \end{threeparttable} 
    \label{tab:cost-estimates}
\end{wraptable}

    \subsection{Cost Model Overview}
\label{sec:cost-model-description}
\label{sec:malware-costs}
This model estimates and sums labor and resource costs incurred for buying, configuring, and using the tool in  
addition to attack damage cost estimates. These costs are itemized by the following components. Refer to Table \ref{tab:cost-estimates}.

\textbf{$C_I$, the one-time initial costs of purchases, setup, and installation.} 
The initial configuration costs are based on our team’s experience with installing and configuring these devices at the NCR. All costs are slightly obfuscated to preserve anonymity of the tool vendors. 
This minimally impacts results but does not impact the rankings of the tools. All license, hardware, and labor costs assume a relatively small network ($\smalltilde$1K IPs), comparable in size and complexity to the NCR test network described previously. Note that we ignore several smaller costs, such  as power and cooling, because they are negligible.

\textbf{$C_B$, the ongoing (per month, year, etc.) costs of normal operation,} including subscription-based licenses and labor needed for periodic updates, reconfiguration, and maintenance. Since we did not use the tools in a real setting for a long period of time, we simply cannot give per-tool estimates. We assumed 8 h/tool/month for all tools. This does not affect our comparison  and was included for completeness.

\textbf{$C_{IR}$, the incident response (IR) cost (per true positive alert),} representing labor costs for investigation and remediation of detected attacks. 
We estimate this at \$280 = \$70/h (fully burdened cost of SOC operator) $\times$ 4 h.

\textbf{$C_T$, the alert triage costs (per alert),} representing labor costs and storage costs for the alert’s data. We
\begin{wraptable}[]{r}{.33\textwidth}
    \centering
    \caption{Cost model parameters}
    \label{tab:cost-model-params}
    \begin{threeparttable}
    \begin{tabular}{ll}
        \toprule
          \textbf{Param. Value} & \textbf{Description} \\
        \midrule
             $M$ =    \$2,000  &        Max. Attack Cost \\
             \rowcolor[HTML]{EFEFEF} 
             $\alpha$ =  15m = 900s & Half-Cost Time \\
              $N$ = 125K& \# of Files\\
             \rowcolor[HTML]{EFEFEF} 
             $r$ = 1.16\%& Malware Ratio\\
        \bottomrule
    \end{tabular}
    \end{threeparttable}
\end{wraptable} 
estimate this at \$70/h $\times$ 1 h. 
A SOC lead verified that these estimates of labor time for triage and (above for) IR were verified as reasonable. 
The triage resource cost of \$0.05 was obtained by finding the lowest volume Splunk license based on the rate of raw data being processed (\$1,800 for 1 GB/month, or 34 MB/day) and estimating what fraction of that volume would  be consumed by each alert. 
For both detectors, we observed a peak of 1,000 alerts per day during our testing, and the average alert size was under 1 KB per alert, for a maximum of 1 MB/day. Because each detector is using less than 1/34th of the licensed volume for its 1,000 alerts per day, each alert represents a cost of (at most) \$1,800/34/1,000 or about \$0.05.

\begin{wrapfigure}[]{r}{.5\textwidth}
    \centering    
    \includegraphics[width=.5\textwidth]{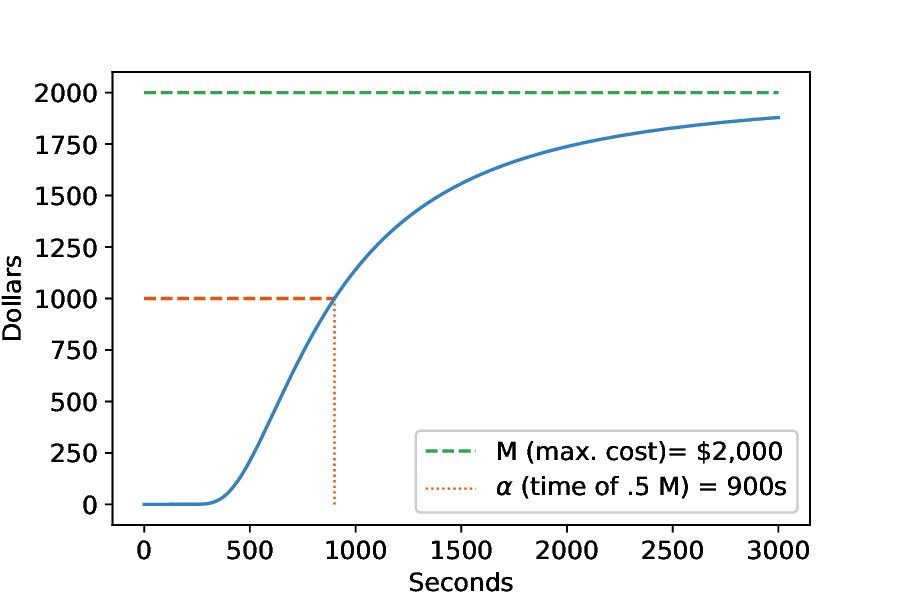}
    \centering
    \caption{Attack cost model, \small{$f(t) = M  2^{-1/(t/\alpha)^2}$}}
    \label{fig:attack-cost}
\end{wrapfigure}

$C_A$: While the above four costs are ``defense costs,'' $C_A$ \textbf{estimates the losses due to attack damages.} Attack cost is a function  of time modeling the average cost incurred from the moment of infection. 
\hl{We follow formulation of attack cost from the initial work {\cite{iannacone2020quantifiable}} and  use an ``S'' curve to model average attack cost over time. The intuition is that costs will begin at \$0 and after some time (e.g., the malware unpacking itself) begin to grow quickly, then level out approaching a maximum cost. 
In reality, this might happen in a sequence of discrete jumps, but as discussed in the previous work, an ``S'' curve provides a reasonable model of the average cost for an attack.}
Specifically, the attack damage cost $f(t)$ is defined as:
$f(t) = M \exp(-  (\alpha/t)^2 \ln{2}) = M  2^{-1/(t/\alpha)^2}$
where parameter $M$ is the maximum cost (and horizontal asymptote), and parameter $\alpha$ is the time when half the maximum cost is attained ($f(\alpha) = M/2)$). 
See depiction of $f(t)$ in Figure \ref{fig:attack-cost} and input parameter choices 
in Table \ref{tab:cost-model-params}. 
The shape of the attack cost function, the parameters in Table \ref{tab:cost-model-params}, and the cost of labor hours used in this model were informed by discussions with a SOC lead who verified they are reasonable. 

To apply the model we first computed initial and ongoing costs, as explained previously, then compute each tool's cost per sample as follows:
\begin{description}
\item[False Negative Costs: ] If the tool does not alert on a malicious sample, maximum attack cost is incurred ($M = \lim_{t\to\infty} f(t)$).
\item[True Positive Costs:] If the tool correctly detects malware in $t$ seconds, the incurred cost is the sum of costs from triage resource and labor ($\$0.05 + \$70$), IR labor ($\$280$), and attack damages $f(t)$. 
Very late detection of malware can cost more than no detection because, for large $t$, the attack cost is near maximal, and alert triage and remediation still occurs.  
\item[False Positive Costs:] If the tool incorrectly alerts on a benign sample, triage costs are incurred ($\$0.05 + \$70$).
\item[True Negative Costs: ] If the tool (correctly) does not alert on benignware, no extra costs are incurred. 
\end{description}
Note that FN and TP costs are costs from malicious samples, whereas FP and TN costs are from benign samples.

\hl{We break up the sum of $C_{IR} + C_T + C_A$ (incident response, triage, and attack costs) into two quantities---costs from false alerts plus from attacks (true alerts or no alert). 
To do so, we compute each tool's average benignware cost $a_b$  (from the remaining 992 benign samples) and average malware cost $a_m$ (from the 2,554 malware samples). 
Finally, we set $N$, the number of unique files crossing the network detector (border traffic + internal-to-internal) for a 1,000 IP network, to be 125K files (125 per host).
We also need an estimate of the ratio of malware to benignware in the wild. For this we leverage Li et al. {\cite{li2017large}}, who upon investigating files on 10K IPs find 1.16\% malware and 98.84\% benignware.
} 
Thus, we can scale the average cost per benign 
or malicious file to 125K files per year, respecting the 98.84/1.16\% benign-/malware ratio. 

\hl{Putting it all together, the \textit{estimated cost for the first $y$ years is:} }
\begin{align} 
\label{eqn:cost}
    C   &= C_I      &&\text{initial resource and labor costs plus} \nonumber \\
        &+ y[  &&\text{years ($\times$ quantity)} \nonumber  \\
        &  \phantom{+ y[ + } C_B && \text{yearly ongoing costs, e.g., subscriptions and tweaking, plus} \\
        &   \phantom{+ y[ } + N \times 98.84\% \times a_b && \text{annual costs from false alerts, plus} \nonumber \\
        &  \phantom{+ y[ } + N \times 1.16\% \times a_m] && \text{annual costs from attacks (end quantity)} \nonumber 
\end{align}

\begin{figure*} 
\centering    
    \includegraphics[width=.48\linewidth]{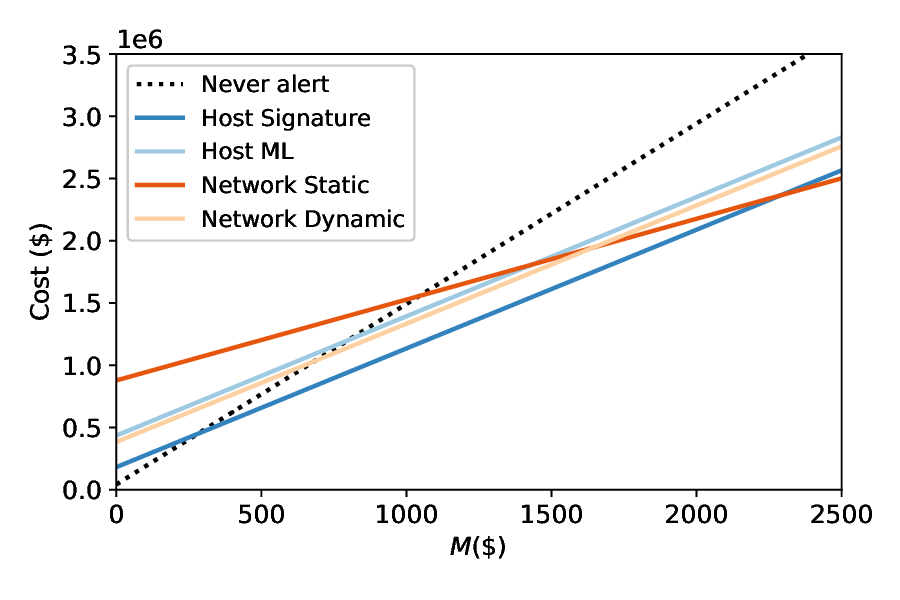}
    \includegraphics[width=.48\linewidth]{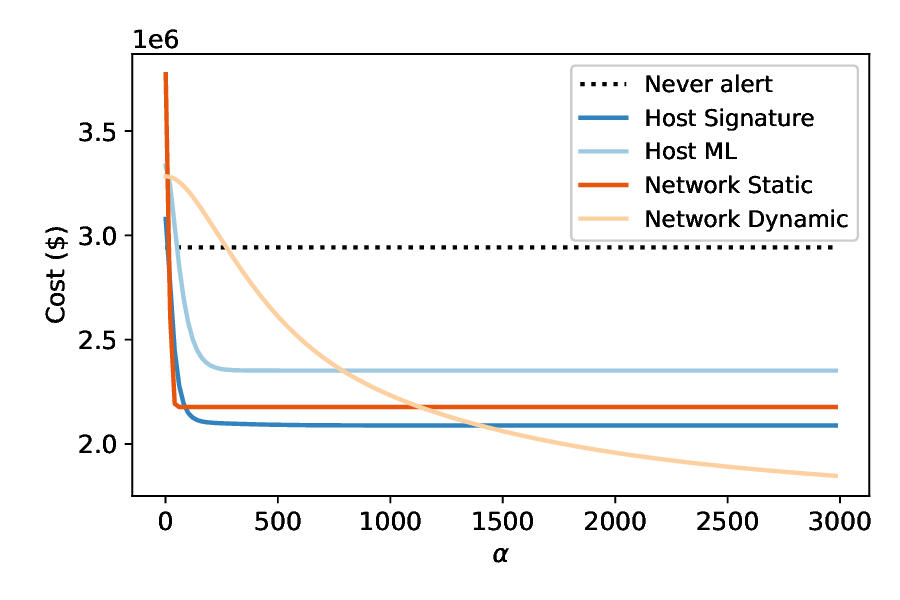}
    \caption{Cost model results for varying $M = $ maximum attack cost with half-cost time $\alpha = 15$m (left); varying $\alpha$ with $M =\$2000$  (right).  Regarding the \hl{left} plot,  for $M$ near 0, it is cheaper to never detect (never alert baseline is cheapest). For $M \in [\smalltilde\$300, \smalltilde\$2300]$ the host signature detector is best and is second to the network static tool for $M>\smalltilde\$2300$.  Regarding the \hl{right} plot, for small (large) $\alpha$, attack cost accrues quickly (slowly), with the effect that slow  detection will be very (not very) penalized. The lone dynamic tool is very slow to detect but very accurate; hence, with large $\alpha$ it dominates. Of the two host-based tools, the signature-based method wins for nearly all parameters.}
    \label{fig:vary-params}
\end{figure*}

Importantly, $N$ linearly scales all costs except the initial and ongoing costs (see in Figure \ref{fig:vary-params} (Left)), so unless the initial and ongoing costs for a tool are exorbitant, inaccuracies in this file number estimate will not affect any comparison of the tools.
For the latter two, we assumed detection was ``immediate'' using a detection time of $t = 1\mathrm{E}{-10}$~s. 





\subsection{Single-Tool Cost Model Results}
\label{sec:cost-model-results}\label{sec:result-one-tool}
For comparison in the subsequent results, we include three simulated  detector devices: ``Never Alert,'' which simulates no detector; ``Always Alert,'' which alerts on every file; and a ``Perfect Detector,'' which correctly alerts on only malware.
\begin{wraptable}{r}{.5\textwidth}
    \centering    
\caption{Cost model results---mean cost per benign-/malware ($a_b$ , $a_m$) and 1st year cost---with three simulated baselines using parameters as in Table \ref{tab:cost-model-params}. Cost increases linearly in per-file costs. Mean benignware cost is a multiple of the false alert rate, while resp. mean malware cost is monotonically related to the recall and time to detect via attack cost function.}
\label{tab:cost-results}
\begin{threeparttable}
\begin{tabular}{lccc}
\toprule
{} &  Benign & Malware & first  1st Year\\
\cmidrule{2-4}
Host Sig.       &            \$0.00 &          \$1,435.47 & \$2,098,712\\
\rowcolor[HTML]{EFEFEF}
Host ML               &             \$2.07 &          \$1,440.54 & \$2,393,151\\
Network Dyn.      &            \$0.50 &          \$1,500.41 & \$2,283,562\\
\rowcolor[HTML]{EFEFEF}
Network Static   &               \$4.49 &          \$1,089.10 & \$2,176,719\\
\midrule
Never Alert             &             \$0.00 &          \$2,000.00 & \$2,942,280 \\
\rowcolor[HTML]{EFEFEF}
Always Alert &            \$70.05 &            \$350.05 & \$9,204,530 \\
Perfect Detector              &             \$0.00 &            \$350.05 &   \$549,852 \\
\bottomrule
\end{tabular}
\end{threeparttable}
\end{wraptable}

Table \ref{tab:cost-results} displays the average benign-/malware costs and the estimated cost of using each tool alongside the simulated baseline under parameters as shown in Table \ref{tab:cost-model-params}. 
First and foremost, the simulations provide validation of the model, as all tools perform much better than the always alert simulation (accruing enormous false positive cost but minimal true positive cost) and much worse than the perfect detector device.
Because malware is relatively rare (ratio of 1.16\%), the cost of never alerting, though still more than all actual detectors, is closer than the other two simulated baselines. 
The tool rankings show that the host signature tool is superior because it has no FPs and competitive time to detect. 
As we only consider the host tools for realistic standalone defenses, we note that even with identical initial and ongoing costs, the host \hl{signature-based} tool is superior.
Interestingly, the network static detector places second because it is by far the fastest detector and has high recall, both outweighing its higher FP rate (high Ave \$/Benign). 
The network dynamic tool is hindered by its slow detection time (Figure \ref{fig:tc2:malware-analysis-times}) but benefits from accuracy. 

\subsubsection{Insights from Varying Cost Model Parameters}
\label{sec:result-vary-params}
The skeptic might object to  cost-benefit analyses, citing that, especially for cybersecurity applications, such methods suffer from required inputs that are fundamentally difficult to ascertain \cite{iannacone2020quantifiable, butler2002security};  and, indeed, our model is no different. 
To account for these unknowns, this section investigates parameter sensitivity.


Costs for all tools scale linearly in $N$, the number of files, so this will not change rankings. 
Cost increases linearly with mean \$/benignware with weight $N\times (1-r)$ and mean \$/malware with weight $r$, where $r$ is the malware ratio. 
Ave\$/benign is a multiple of the false alert rate,  while Ave\$/malware is monotonically related to the tool's recall and time to detect via attack cost. 
Thus, increasing/decreasing $r$ decreases/increases impact of the false alert rate but increases/decreases the impact of recall and time to detect. 
Precision directly affects results, but recall's effect depends indirectly on $M$ and $\alpha$. 
To investigate, we present costs while varying $M$ and $\alpha$ in Figure \ref{fig:vary-params}. 
The main takeaway is that the host signature tool is the best standalone choice based on these results and cost model. 
It offers the lowest cost for nearly all values of $M$ and $\alpha$, and otherwise (Figure \ref{fig:vary-params}) is second to network tools. 

\begin{figure*} 
\centering    
    \includegraphics[width=.48\linewidth]{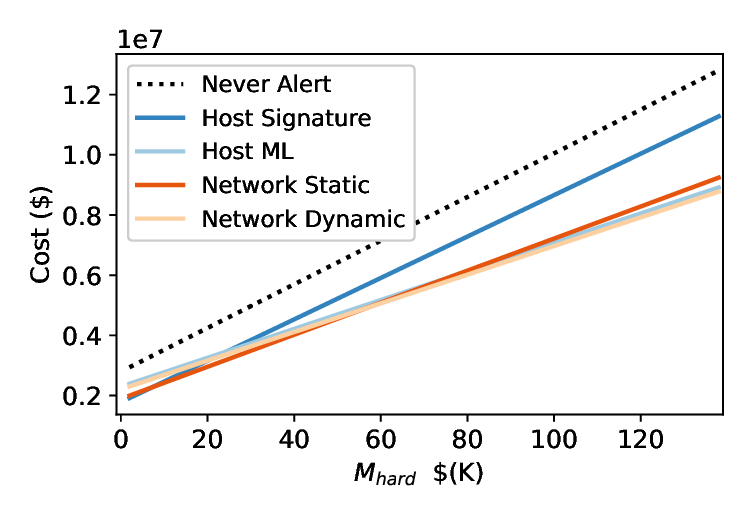}
    \includegraphics[width=.48\linewidth]{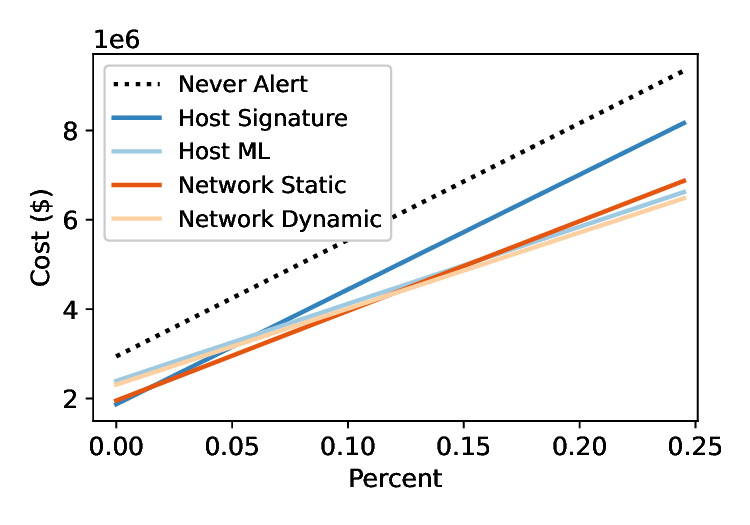}
    \caption{\hl{Cost model results for varying $M_{hard} = $ maximum attack cost for hard files (namely polyglot, zero-days, and APT-style files) with percent of hard malware $p=0.05$ (left) and varying $p$ the percent of malware that is hard with $M_{hard} = \$20,000$ (right). 
    In both, the same half-cost time $\alpha = 15$ m is used for all malware, and for easy malware, $M =\$2,000$. 
    First notice that both plots are similar in structure because of the linear nature of these variables---both increases in $M_{hard}$ and $p$ convert savings from those with low recall  to those with better recall on hard files. 
    Hence, in both plots, we see that for small values of $M_{hard}$ and $p$ the signature-based detector wins; for intermediate values, the network static (relatively faster detection, competitive recall) wins; yet for larger values, the host ML and network dynamic (best recall on hard files) win. These results are intuitive, which is reassuring, but force a difficult reality; that is, deciding which tool is best depends on one's estimate of the hard'' malware expected. This suggests pairing signature- and ML-based tools. }}
    \label{fig:vary-params-hard}
\end{figure*}
\subsubsection{Insights from Increasing Cost of ``Hard'' Files---Zero-days, Polyglots, and APT-style Files} 
Another addition to the original cost model that is applicable to this study is increasing the cost of the ``hard'' malware: zero-days, polyglots, and APT style files, representing the assumption that such files, if undetected,  will result in more damage than $n-$day files, on average. 
To exhibit results under these conditions,  we compute  $a_{mh}$ the average hard file cost using the detection times on only the zero-days, polyglots, and APT-style functions and with augmented maximum attack cost $M_{hard}$ (still using the same equation with identical $\alpha$).
Similarly, we compute $a_{me}$, the average ``easy'' file cost identically on the remaining ($n-$day) malware detection results and using a maximum attack cost of $M = \$2K$.
Finally, we replace the final line of Equation \ref{eqn:cost} (namely, $N \times 1.16\% \times a_m$ ) with a convex sum of  the average cost from hard malware files and from easy malware files (namely, $N\times 1.16\% \times p  \times a_{mh} + N\times 1.16\% \times (1- p)\times a_{mh}$) using a parameter $p$, the percent of malware assumed to be hard. 
Results are depicted in Figure \ref{fig:vary-params-hard}. 
The main takeaway is that if one expects (1) more than a couple percent of malicious files to be a zero-day, polyglot, or APT-style file and (2)  the cost from an undetected hard file to be much greater than otherwise, ML-based detectors provide substantial savings.

\subsection{Savings by Adding Network Detectors}
\label{sec:savings-addition}
Here we describe a new way to configure the cost model that is likely more useful for real-world use---estimating the cost/savings of adding each network malware detector to each host-based detector. 
To do so, we add the initial and ongoing costs for the network tool then compute the difference in cost between using both the host-based and network-based tools and solely the host-based tool.  
As before, we (1) compute the additional cost/savings per file, (2) find the average cost/savings per malware file and per benignware file, and (3) linearly scale these costs/savings to 125K total files, respecting the 98.84/1.16\% benign-to-malicious file ratio. 
To estimate the change in cost on a given file, if the network   tool  does not alert on a given sample, 
there is no change in cost. 
\begin{wraptable}[10]{r}{.66\textwidth}
\centering
\caption{Cost on average per file and for first year of network detection tools  when added to each host-based detector individually. Negative costs indicate savings.}
\label{tab:cost-add-to-host}
\begin{threeparttable}
\begin{tabular}{lcrccrcc}
\toprule
&&& \multicolumn{2}{c}{\textbf{Host ML}} & & \multicolumn{2}{c}{\textbf{Host Signature}} \\
 \cmidrule{4-5} \cmidrule{7-8}
{} &  Benign &&  Malware &  1st Year  &&    Malware &  1st Year \\
\cmidrule{2-2} \cmidrule{4-5} \cmidrule{7-8}
\textbf{Dyn.}   &               \$0.50 && \$-145.26 &         \textbf{\$-102,651}
&&                          \$-298.22 &               \textbf{ \$-324,443}
\\
\rowcolor[HTML]{EFEFEF}
\textbf{Stat.} &               \$4.49 &&         \$-434.09 &          \$-31,913
&&                          \$-427.36 &                 \$-22,148\\
\bottomrule
\end{tabular}
\end{threeparttable}
\end{wraptable} 
If the network tool alerts on a sample (FP or TP), then triage resource and labor costs ($\$0.05 + \$70$) are incurred, accounting for an operator fielding the alert from the network tool (possibly in addition to alerts from the host-based detector). 
Moreover, if this network-based alert is a TP, and the host-based tool also correctly alerts on the sample, we make no other change to costs for this file; only triage cost is added for the network alert. 
This follows two assumptions: (1) IR actions will have taken place with or without the network-based alert since the host-based tool correctly alerted and (2) host-based detection is at least as fast as the network detector, resulting in no change in attack cost. 
These assumptions are based on the fact that most host-based tools offer \textit{pre-execution file conviction} and \textit{blocking}. 
As the final case, consider when the network tool's alert is a TP but the host-based tool failed to alert on the malware. 
In this scenario, without the network-based tool, the attack goes undetected, and the maximum attack cost is incurred. Yet with the network-based tool, there are now triage and IR costs, but the attack cost is reduced from maximal to $f(t)$, the attack cost at $t = $ the network-level detection time. 
In our model, this is the lone scenario in which savings from the network-level tool can occur---when the network tool identifies malware that is undetected by the host tool and the network tool identifies it fast enough that the change in attack cost is lower than the incurred triage and IR costs.

Table \ref{tab:cost-add-to-host} reports the results of adding both  network-based tools to the two host-based detectors with parameters as in Table \ref{tab:cost-model-params}.  
Reassuringly, both network-level detectors yield substantial (\$22K--\$325K) savings when added to the host tools. 
These results predict that, for SOCs using either host detector, dynamic detection is the best pairing, saving  3--15$\times $ more than the static detector.

\section{Conclusions, Limitations, Next Steps, and Takeaways}
\label{sec:did-discussion}
This study describes experiments with four \hl{prominent} malware detection technologies aimed at helping an organization assess (Q1) the ML generalization hypothesis and (Q2) the added value of network-level malware detectors. 
Our results provide empirical quantification on the efficacy of four market-leading  malware detectors. 
The three ML tools demonstrated a minimal 2$\times $ increase in detection coverage for publicly available executables and a 10$\times $ increase in malware detection coverage for zero-day executables.  
Less intuitive perhaps was how well the host signature--based tool performed:  exhibiting the best (perfect) precision; demonstrating comparable recall ($\smalltilde$35\%) to the host ML--host tool; and, according to the cost model, performing the best overall \hl{for small percentages of hard (zero-day, polyglot, or APT-style) files and when maximum possible damage cost incurred by an undetected hard file was not much more than an $n-$day malware.
Notably, if one expects (1) more than a couple percent of malicious files to be zero-day, polyglot, or APT-style files, and (2)  the cost from an undetected hard file to be much greater than otherwise, ML-based detectors provide substantial savings; this provides a strong argument for pairing detectors.} 
Combining the signature-based tool with an ML-based tool enables malware detection ranges above 90\% for in-the-wild executables and above 60\% for zero-day malware. 
As demonstrated by increased recall and the new configuration of the cost model, substantial value is added by combining either of the two network-level tools to either host-level tool. Unsurprisingly, dynamic analysis incurs the cost of latency for increased recall, and both network-level detectors vary in their capabilities across protocols. 
Our results indicate that low false positive rates (precision $> $98\%) should be expected, with fast detection time for static detectors, but that this comes at the cost of recall. 
Of the four detectors in this study, none surpassed 55\% recall, and recall falls dramatically with JAR and polyglot file types.

Testing only four representative tools is a clear limitation; in particular, 
many COT tools offer (and some require) cloud connectivity to enhance detection at the expense of privacy, but no such tool was represented. 
Our test environment did not allow connection to the Internet, which might have stymied malware actions and therefore affected detection results. 
\hl{Similarly, network-level signature-based tools are not fully represented. }
Impact to host resources was not taken into account in this study (CPU, memory, and disk I/O, in particular). 
More sophisticated (and presumably more accurate) models of attack cost could be integrated in future work (e.g., \cite{Kondakci_2009}). 
Finally, accurate cost estimates from the cost model will require per-SOC inputs. Based on our parameter sensitivity analysis, the rankings and takeaways of the cost model results for these tools should generalize (and our treatment provides a blueprint for future use). 
Nevertheless, we argue that trends found among these four market leaders are sufficient to  pose hypotheses, raise awareness of gaps, and sharpen next-step research.

We itemize priorities for next-step research from this work: 
Our polyglot detection results suggest that detectors are failing at preclassifying a file into possibly many file types for which it can execute. 
Research is needed to accurately and quickly determine \textit{all} of the file types of a given sample and integrate this capability into COTS detectors to enhance detection. 
Further research is needed to address the polyglot detection issue and performance with throughput and computational expense in mind. 
More accurate detection models for all file types are needed. 
In particular, future malware detection research should focus on increasing the detection capabilities for non-PE and non-MS Office file types. 
This might entail enhancing the quantity of malware and benignware files of many types that are currently less prevalent (e.g., through an effort to build and make public a more diverse malware data set). 
For the two network-level detectors, recall varied widely when identical files were delivered on different protocols. 
We suggest a study to identify whether the file carving is the limiting factor and to find methods for accurately extracting a copy of the file from the packet stream to the detector. 
\hl{Future work to tune the cost model to a specific SOC is both reasonable and likely very valuable if successful, especially if it produces a procedure and code-base for refitting to any SOC. 
Malware detection studies in which computational resources of the hosts are monitored and incorporated are needed, in particular for a higher fidelity cost model.  
As evidenced by our results, many of the takeaways depend on difficult assumptions; for example, the percent of files that are malware (e.g., Li et al. {\cite{li2017large}}) or the percent of different file types or categories in the wild. Empirical studies to provide data on cyber-related artifacts provides fodder for useful studies. }

Finally, some takeaways for SOCs. 
Great gains in detection rate (recall) with few false positives (high precision) are permitted by all pairs of tools in our study, but little is gained from using more than two of these tools. 
When choosing a pair of tools, strive for tools that complement each other in terms of increasing the coverage (detection rate) of all file types, rather than duplicating results---this might require some preliminary testing with different file types. 
A feasibility study for any desired pairing of host tools is needed before deployment to examine how the combination will affect the host’s resources. 
Our results (Table \ref{tab:cost-add-to-host}) show that substantial savings are yielded by adding a network-level detector in all four combinations of network-to-host detectors tested.
The accuracy gains permitted by the dynamic detector in our study outweighed latency costs. 
For any network detector, testing bandwidth and file rates with the specific software and hardware is needed. 
Note that network-level detectors require unencrypted traffic, so we are presupposing that an SSL/TSL ``break and inspect'' technology is in place.

\begin{acks}
Thanks to Charlie Horak for the editorial review, and to the many reviewers whose constructive criticism helped polish this work. 

The research is based upon work supported by the Department of Defense (DOD), Naval Information Warfare Systems Command (NAVWAR), via the Department of Energy (DOE) under contract  DE-AC05-00OR22725. The views and conclusions contained herein are those of the authors and should not be interpreted as representing the official policies or endorsements, either expressed or implied, of the DOD, NAVWAR, or the U.S. Government. The U.S. Government is authorized to reproduce and distribute reprints for Governmental purposes notwithstanding any copyright annotation thereon. 
\end{acks}

\bibliographystyle{ACM-Reference-Format}
\bibliography{refs}

\break

\section*{Appendix}
\appendix
\begin{figure}[!h]
    \centering
    \includegraphics[width=\textwidth]{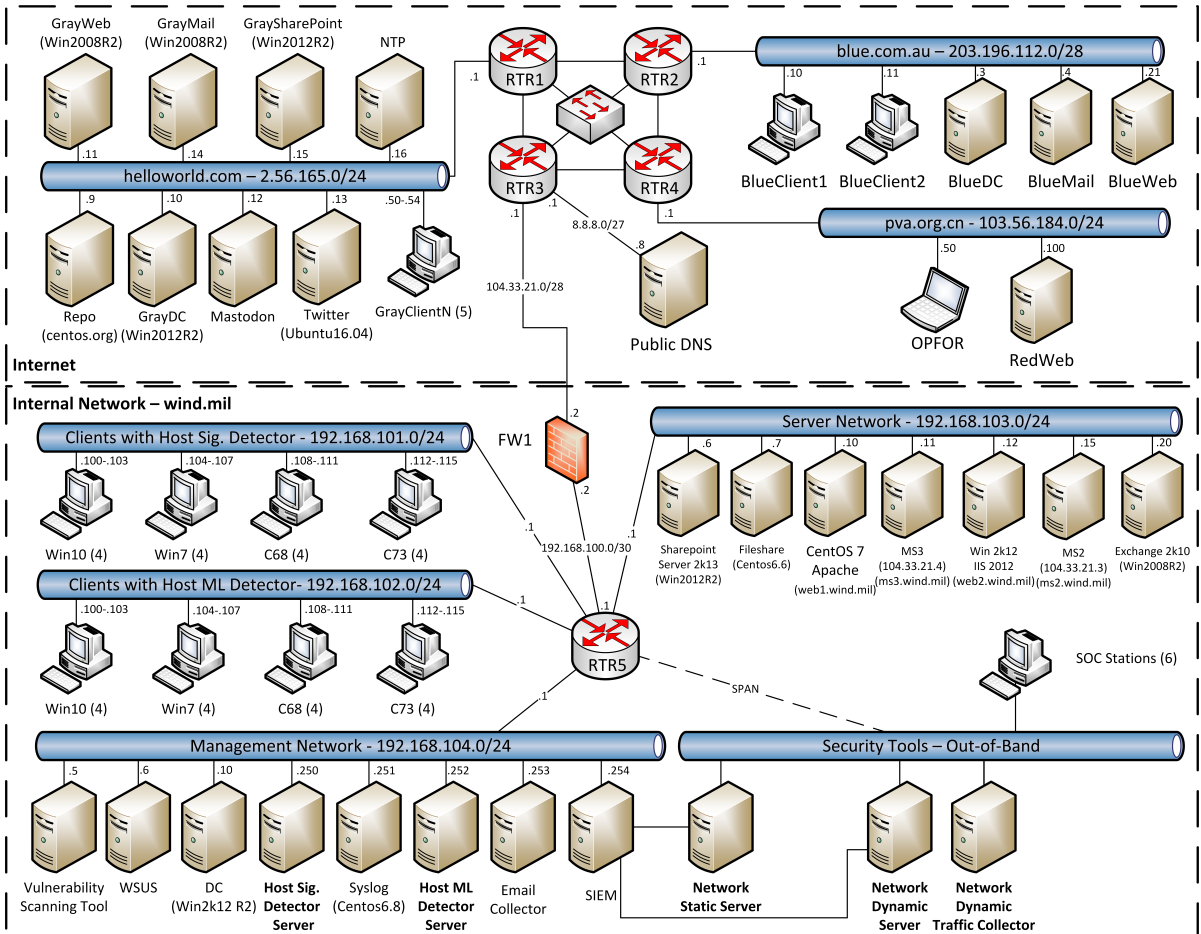}
    \caption{\hl{Network Diagram}}
    \label{fig:network-diagram}
\end{figure}

\end{document}